\documentclass[11pt]{article}
\usepackage{amssymb}
\usepackage{amsmath}
\usepackage{amscd}
\usepackage{latexsym}

\catcode`@=11
\renewcommand{\section}{\@startsection{section}{1}{0pt}{\medskipamount}
{\medskipamount}{\large\bf}}
\numberwithin{equation}{section}
\catcode`@=12
\def\a{\alpha}
\def\b{\beta}
\def\g{\gamma}
\def\de{\delta}
\def\eps{\epsilon}
\def\ve{\varepsilon}

\def\h{\eta}
\def\th{\theta}

\def\la{\lambda}
\def\m{\mu}
\def\n{\nu}
\def\r{\rho}
\def\s{\sigma}
\def\p{\phi}
\def\vp{\varphi}

\def\Ups{\Upsilon}

\newcommand{\C}{\mathbb C}
\newcommand{\R}{\mathbb R}
\newcommand{\Z}{\mathbb Z}
\newcommand{\N}{\mathbb N}
\newcommand{\Acal}{{\cal A}}
\newcommand{\Ecal}{{\cal E}}
\newcommand{\Hcal}{{\cal H}}
\newcommand{\Lcal}{{\cal L}}
\newcommand{\Ocal}{{\cal O}}

\newcommand{\Zcal}{{\cal Z}}
\newcommand{\U}{{\cal U}}

\def\e{\mbox{e}}
\def\im{\mbox{i}}
\def\N2{$N{=}2$}
\def\pa{\mbox{$\partial$}}
\def\diff{\mbox{d}}
\def\tr{{\rm tr}}
\def\sfrac#1#2{{\textstyle\frac{#1}{#2}}}

\newcommand{\fh}{\hat{f}}
\newcommand{\xh}{\hat{x}}

\newcommand{\wh}{\hat{w}}
\newcommand{\yh}{\hat{y}}
\newcommand{\zh}{\hat{z}}
\newcommand{\rh}{\hat{\rho}}
\newcommand{\ph}{\hat{\psi}}

\newcommand{\lb}{\bar{\lambda}}
\newcommand{\yb}{\bar{y}}
\newcommand{\zb}{\bar{z}}

\newcommand{\lt}{\tilde{\lambda}}
\newcommand{\te}{\tilde{\eta}}
\newcommand{\tw}{\tilde{w}}
\newcommand{\rt}{\tilde{\rho}}
\newcommand{\cp}{\C P^1_x}

\newcommand{\va}{\Vec{a}}

\def\>{\rangle}
\def\<{\langle}
\def\+{\dagger}
\def\={\ =\ }

\begin{document}
\begin{titlepage}
\setcounter{page}{0}
\begin{flushleft}
\end{flushleft}
\begin{flushright}
hep-th/0306263\\
ITP--UH-04/03\\
\end{flushright}

\vskip 2.0cm

\begin{center}

{\Large\bf  
Noncommutative Monopoles and Riemann-Hilbert Problems
}

\vspace{14mm}

{\Large
Olaf Lechtenfeld} \ \ and \ \
{\Large
Alexander D. Popov~$^{*}$}
\\[5mm]
{ \em
Institut f\"ur Theoretische Physik  \\
Universit\"at Hannover \\
Appelstra\ss{}e 2, 30167 Hannover, Germany }
\\[5mm]
{Email: lechtenf, popov@itp.uni-hannover.de}

\end{center}

\vspace{15mm}

\begin{abstract}
\noindent
The Bogomolny equations for Yang-Mills-Higgs monopoles follow from a system
of linear equations which may be solved through a parametric Riemann-Hilbert
problem. We extend this approach to noncommutative $\R^3$ and use it to
(re)construct noncommutative Dirac, Wu-Yang, and BPS monopole configurations
in a unified manner. In all cases we write down the underlying matrix-valued
functions for multi-monopoles and solve the corresponding Riemann-Hilbert
problems for charge one.
\end{abstract}

\vfill

\textwidth 6.5truein
\hrule width 5.cm
\vskip.1in

{\small \noindent 
${}^*$
On leave from Bogoliubov Laboratory of Theoretical Physics, JINR,
Dubna, Russia
}

\end{titlepage}

\section{Introduction } 

\noindent
In modern string theory, solitonic solutions to the (effective) field equations
are ubiquitous. A particular class of solitons are monopoles~\cite{Witten}, 
which appear as D1-branes stretched between D3-branes in type~IIB superstring 
theory. When a constant NS $B$-field is turned on, gauge theory on a stack of 
D3-branes becomes noncommutative and supports noncommutative U$(n)$ monopoles, 
including abelian ones~\cite{Hashimoto:1999zw}\,--\,\cite{Hamanaka:2003cm}.

Irrespective of string theory, monopoles are playing an important role in the
nonperturbative physics of $3{+}1$-dimensional Yang-Mills-Higgs theory
\cite{'tHooft}\,--\,\cite{Rajaraman}. 
In the Prasad-Sommerfield limit interesting monopole solutions arise from a
Bogomolny argument on the energy functional~\cite{Bogomolny}.
The simplest such configurations can be given in explicit form~\cite{Prasad1}, 
due to the integrability of the Bogomolny equations, which in fact emerge via 
reduction of the self-dual Yang-Mills equations from four to three dimensions.
More specifically, the Bogomolny equations may be written as the compatibility
condition of an auxiliary linear system of differential equations whose 
solutions in turn can be obtained from a parametric Riemann-Hilbert problem.
Mathematically, the latter is most naturally formulated in minitwistor space
\cite{Hitchin}.

The computational task then is, firstly, to make an educated guess for 
a matrix-valued function~$f$ of the coordinates $(x^1,x^2,x^3)\in\R^3$ and of 
the spectral parameter~$\la\in S^1\subset\C P^1$ and, secondly, 
to factorize~$f=\psi_+^{-1}\psi_-$, with $\psi_+$ and $\psi_-$ extendable
holomorphically onto complementary patches of~$\C P^1$. This is sometimes
called the ``splitting method''~\cite{Ward-Wells}. Finally, the gauge 
potentials are straightforwardly recovered from the matrices~$\psi_{\pm}$.
In this fashion, the generic Bogomolny-Prasad-Sommerfield (BPS) monopoles have
been constructed a long time ago, albeit only for charges~$|Q|\le2$ explicitly
\cite{Ward1,Ward2,Corrigan2,Forgacs}.
We successfully apply this technique also to the abelian Dirac as well as the 
point-like nonabelian Wu-Yang monopoles.

The three known methods for generating solutions to the Bogomolny equations
-- splitting, dressing, and Nahm's approach -- all build on the fact that
these equations are the compatibility conditions of some Lax pair.
Fortunately, noncommutative deformation respects this integrable structure.
Therefore, the just-mentioned solution-generating techniques can be generalized
to the noncommutative setup. Indeed, Gross and Nekrasov~\cite{Gross:2000wc}
directly deformed the Nahm equations~\cite{Nahm1} and constructed a nonsingular
abelian monopole on noncommutative~$\R^3$. Along the same lines, noncommutative 
nonabelian BPS monopoles were found~\cite{Gross:2000ss}. 
Their commutative limit reproduces the ``ordinary'' monopoles.

The above-mentioned splitting method provides an alternative approach.
It has been adapted to the noncommutative situation and exploited successfully 
for instanton construction~\cite{LP1,HLW}.
In the present paper we apply the (deformed) splitting method to a variety of
monopoles, first commutative and then noncommutative.  After briefly 
reviewing Yang-Mills-Higgs theory and its static monopole configurations,
including the noncommutative extension, in section~2, we present in section~3
the road from the Bogomolny equations to the Riemann-Hilbert problem and back
to the monopole configuration in some detail. Sections 4, 5, and~6 are devoted
to Dirac, Wu-Yang, and BPS monopoles, in that order, where the commutative
construction precedes the noncommutative one in each case. It is remarkable
that all three kinds of monopoles are parametrized by the same function
appearing in their respective Riemann-Hilbert problems. Also, the derivation
of noncommutative multi-monopoles is outlined, but performed only for the
Wu-Yang case. For the more mathematically inclined reader, three appendices 
explain the twistor reformulation and its noncommutative extension,
both for self-dual gauge fields and for monopoles, and demonstrate the
equivalence of the general Wu-Yang and Dirac multi-monopoles.

\newpage

\section{Yang-Mills-Higgs model in Minkowski space} 
 
\subsection{'t~Hooft-Polyakov monopoles} 

\noindent 
{\bf Notation.} 
We consider the Minkowski space $\R^{3,1}$ with the metric  
$(\h_{\m\n})={\rm diag}(-1,1,1,1)$, a real scalar (Higgs) field $\p$, a gauge 
potential $A{=}A_\m dx^\m$ and the Yang-Mills field $F{=}\diff A{+}A{\wedge}A$ 
with components $F_{\m\n}=\pa_\m A_\n -\pa_\n A_\m + [A_\m , A_\n ]$, where  
$\pa_\m :=\pa /\pa x^\m$ and $\m , \n ,\ldots = 0,1,2,3$. The fields $A$ and $F$
take values in the Lie algebra $u(n)$. The matter field $\p$ belongs to the 
adjoint representation of the group U$(n)$ for any $n=1,2,\ldots\ $. We choose 
normalization of the generators $T_i$ of this group such that $\tr (T_iT_j)= 
-\de_{ij}$ for any representation, $i,j,\ldots=1,\ldots,n$. 
 
\noindent 
{\bf Lagrangian.} 
Consider the following Yang-Mills-Higgs Lagrangian 
density~\cite{'tHooft, Polyakov}: 
\begin{equation}\label{ld} 
\Lcal \= - \sfrac{1}{4}\tr (F_{\m\n}F^{\m\n} +2\h^{\m\n} D_\m\p D_\n\p )- 
\sfrac{\g}{4}(|\p |^2-1)^2\ , 
\end{equation} 
where $\ D_\m = \pa_\m + [A_\m ,\ . \ ]\ $, $\ |\p|^2:= -\tr\p^2\ $ and 
$\g\ge 0$ is a real constant. 
The equations of motion following from (\ref{ld}) are 
\begin{equation}\label{em} 
D_\m F^{\m\n} \= -\h^{\n\s}[D_\s\p , \p ]\quad ,\qquad  
\h^{\m\n}D_\m D_\n\p \=\g\p (|\p |^2{-}1)\ . 
\end{equation} 
If the fields $A$ and $\p$ are independent on time $x^0$ and $A_0=0$ then 
the field configuration $(A_a,\p )$ will be called static. 
 
\noindent 
{\bf Energy.} 
For static configurations $(A_a,\p )$ the standard energy functional 
of the system (\ref{ld}) reduces to  
\begin{equation}\label{ef} 
E \= - \sfrac{1}{4}\int\!\diff^3 x\ 
\bigl\{\tr (F_{ab}F_{ab} +2 D_a\p D_a\p ) + {\g}(|\p |^2{-}1)^2 \bigr\}\ , 
\end{equation} 
where $a,b,\ldots=1,2,3$. This functional reaches  a minimum $E=0$ when  
$A_a=0,\ \pa_a\p =0$ and $|\p |^2=1$. For $\g\ne 0$ nonsingular static 
configurations $(A_a,\p )$  with finite energy exist~\cite{'tHooft, 
Polyakov, JT} and are called 't~Hooft-Polyakov monopoles. Finite 
energy ensures that  static solutions to (\ref{em}) satisfy the boundary 
conditions 
\begin{equation}\label{bc} 
\tr (F_{ab}F_{ab})\to 0\quad, \qquad 
\tr (D_a\p\,D_a\p)\to 0\qquad {\rm and}\quad 
|\p |^2\to 1 
\end{equation} 
as $r^2\equiv x^ax^a\to\infty$. 
{}From now on we restrict ourselves to the case of  
U(1) and SU(2) gauge groups, for the general case see e.g.~\cite{JT}.  
 
\noindent 
{\bf Topological charges.} 
Static SU(2) configurations $(A_a,\p )$ are classified  
by the topological charge (number of monopoles) 
\begin{align} 
Q &\=\frac{1}{8\pi}\int\!\diff^3 x\ \eps^{abc}\tr(\pa_a\p\pa_b\p\pa_c\p ) 
\=\frac{1}{8\pi}\int_{S^2_{\infty}}\!\diff^2s_a\ 
\eps^{abc} \tr(\p\pa_b\p\pa_c\p ) \nonumber\\[8pt] 
&\=\frac{1}{8\pi}\int_{S^2_\infty}\!\diff^2\xi\ \eps^{\a\b}\tr 
\Bigl(\p\,\frac{\pa\p}{\pa\xi^\a}\,\frac{\pa\p}{\pa\xi^\b} \Bigr)\ , 
\label{tc} 
\end{align} 
where $\xi^\a$ are coordinates (angles) on the sphere $S^2_\infty$ and
$\eps^{\a\b}:=\eps^{\a\b 3}$ with $\a,\b,\ldots =1,2$. Clearly, (\ref{tc}) 
is the winding number $\pi_2(S^2)=\Z$ for the map $S^2_\infty\to S^2\subset 
su(2)$. 

Note that the definition (\ref{tc}) is equivalent to the following:
\begin{equation}\label{tcSU(2)}
Q \= -\frac{1}{8\pi}\int_{\R^3}\!\diff^3 x\ \eps^{abc}\tr (F_{ab} D_c\p)
\= -\frac{1}{4\pi}\int_{S^2_\infty}\!\diff s_a\ \tr (\sfrac{1}{2}\eps_{abc}
F_{bc}\p )\= \frac{1}{4\pi}\int_{S^2_\infty}\!\diff s_a\ \tilde B_a \ ,
\end{equation}
where $\tilde B_a:= -\tr (\frac{1}{2}\eps_{abc} F_{bc}\p )$ may be identified
with an abelian magnetic field as $r\to\infty$ \cite{'tHooft, JT, Rajaraman}.
In other words, far from its core a nonabelian SU(2) monopole looks like a 
Dirac monopole but with finite energy due to nonsingular nature 
around $x^a{=}0$.
 
In the abelian case the topological charge is defined directly via the flux. 
Namely, the total magnetic flux through a sphere surrounding the origin is 
\begin{equation}\label{mf} 
\Phi \=\int_{S^2}\!\diff s_a\  B_a\ :=\ \int_{S^2}\!\diff s_a\  
\sfrac{\im}{2}\eps_{abc} F_{bc} \= 4\pi Q\ , 
\end{equation} 
where $B_a$ is the magnetic field and $Q$ is an integer. The Dirac monopole 
is defined on the space $\R^3\backslash \{0\}$, which is topologically 
equivalent to $S^2\times\R^+$. The flux (\ref{mf}) calculates the first  
Chern class of  
the U(1) line bundle over $S^2$ and can also be interpreted as the winding  
number $\pi_1(S^1)=\Z$ of the map $S^1\to$ U(1), where $S^1\subset S^2$ 
is an equator in $S^2$ and U(1) is the gauge group. The energy of the  
abelian (Dirac) monopole is infinite. Any superposition of Dirac monopoles  
is a smooth solution on $\R^3\backslash \{\va_1,\ldots,\va_n\}$, where  
$\va_k=(a_k^1,a_k^2,a_k^3)$ for $k=1,\ldots,n$ define the monopole locations.

\subsection{Bogomolny-Prasad-Sommerfield monopoles} 
 
\noindent
Static finite-energy nonabelian solutions of eqs.(\ref{em}) exist for  
any $\g\ge 0$ \cite{JT}, but for $\g\ne 0$ it is not easy to find a solution  
in closed form. The situation simplifies if one puts $\g =0$ 
and keeps the boundary conditions (\ref{bc}). In this case one can derive the  
following (Bogomolny) inequality~\cite{Bogomolny}: 
\begin{align} 
E &\= -\sfrac{1}{4}\int\!\diff^3 x\ \tr (F_{ab}F_{ab} + 2D_a\p D_a\p )
\nonumber\\[8pt] 
& \= -\sfrac{1}{4}\int\!\diff^3 x\ 
\tr (F_{ab} \mp \eps_{abc}D_c\p ) (F_{ab} \mp \eps_{abd}D_d\p ) 
\mp \int\!\diff^3 x\ \tr (\sfrac{1}{2}\eps_{abc} F_{bc}D_c\p ) 
\nonumber\\[8pt] 
& \= -\sfrac{1}{4}\int\!\diff^3 x\ 
\tr (F_{ab} \mp \eps_{abc}D_c\p ) (F_{ab} \mp \eps_{abd}D_d\p ) 
\pm 4\pi Q\ \ge\ 4\pi |Q|\ , 
\label{bog} 
\end{align} 
where $Q\in\Z$ is the magnetic charge (\ref{tcSU(2)}). From (\ref{bog}) it  
follows that the absolute energy minimum $E=4\pi Q$ for a given $Q\ge 0$ 
occurs when the configuration $(A_a, \p )$ satisfies the first order 
Bogomolny equations 
\begin{equation}\label{bog.eq.} 
F_{ab}\=\eps_{abc} D_c\p \ . 
\end{equation} 
{}For $Q<0$, the right hand side of (\ref{bog.eq.}) changes its sign. 
The finite-action solutions of the Bogomolny 
equations are called BPS monopoles since the first explicit $Q{=}1$ 
solution was obtained by Prasad and Sommerfield~\cite{Prasad1}. 
Note that the Bogomolny
equations (\ref{bog.eq.}) are valid also for abelian monopoles but the energy  
will be infinite in this case. 
The main topic of this paper is to demonstrate the power of the splitting 
method for obtaining solutions to the Bogomolny equations and their
noncommutative generalization.

\subsection{Bogomolny equations on noncommutative $\R^3$} 

\noindent
Classical field theory on the noncommutative deformation
$\R^3_\th$ of $\R^3$ may be realized in a star-product formulation
or in an operator formalism.\footnote{
For a review see, e.g. \cite{Douglas:2001ba,Konechny:2001wz}.}
While the first approach alters the product of functions on~$\R^3$
the second one turns these functions~$f$ into linear operators~$\fh$
acting on a Fock space~$\Hcal$.
The noncommutative space~$\R^3_\th$ may then be defined by declaring
its coordinate functions $\xh^1,\xh^2,\xh^3$ to obey
the Heisenberg algebra relations
\begin{equation} \label{ccr1}
[ \xh^a\,,\,\xh^b ] \= \im\,\th^{ab}
\end{equation}
with a constant antisymmetric tensor~$\th^{ab}$.

The coordinates can be chosen in such a way that the only nonvanishing entries
of the matrix $(\th^{\m\n})$ read
\begin{equation}
\th^{12}\= -\th^{21}\ =:\ \th\ \ge0 \ ;
\end{equation}
hence, the coordinate~$\xh^3$ is actually commutative
and we shall retain the notation~$x^3$ for it.
In terms of the complex combinations (to be used later)
\begin{equation}
\yh\= \xh^1+\im\xh^2 \qquad\textrm{and}\qquad \hat\yb \= \xh^1-\im\xh^2
\end{equation}
the basic commutation relations (\ref{ccr1}) become
\begin{equation} \label{ccr2}
[ \yh\,,\,\hat\yb ] \= 2\,\th 
\quad,\qquad \textrm{and all other commutators vanish}\ .
\end{equation}

Clearly, $\yh$ and $\hat\yb$ are (up to a rescaling by $\sqrt{2\th}$)
harmonic-oscillator creation and annihilation operators, respectively.
The corresponding Fock space~$\Hcal$ is spanned by the basis states
\begin{equation}
|k\>\=(2\th k!)^{-\frac12}\,\hat\yb^k\,|0\>
\qquad \textrm{for} \quad k=0,1,2,\ldots \ ,
\end{equation}
so that one has the representation
\begin{equation}
\fh \= \sum_{k,\ell=0}^\infty f_{k\ell}(x^3)\,|k\>\!\<\ell|
\end{equation}
for any linear operator on~$\Hcal$.
We further recall that, in the operator realization $f{\mapsto}\fh$,
derivatives of~$f$ get mapped according to
\begin{equation}
\pa_{y} f \  \mapsto\ -\th^{-1}\,[\hat\yb , \fh] \ =:\ \hat\pa_{y} \fh
\qquad\textrm{and}\qquad
\pa_{\yb} f \ \mapsto\ \th^{-1}\,[\yh , \fh] \ =:\ \hat\pa_{\yb} \fh
\end{equation}
while $\pa_3 f \mapsto \pa_3\fh$.
Finally, we have to replace
\begin{equation}
\int_{\R^3} \!\!\diff^3 x\,f\ \ \mapsto\ \
\int_{\R} \!\!\diff x^3 \ 2\pi\th\,\textrm{Tr}_\Hcal\,\fh\ .
\end{equation}

With this, the noncommutative energy functional (for $\g{=}0$) becomes
\begin{equation}
E_\th \= -\sfrac{\pi\th}{2}\int_{\R} \!\!\diff x^3\ \textrm{Tr}_\Hcal\,
\tr (\hat F_{ab} \hat F_{ab} + 2 \hat D_a\hat\p \hat D_a\hat\p )\ ,
\end{equation}
where, of course, 
\begin{equation}
\hat F_{ab} \= \hat\pa_a \hat A_b - \hat\pa_b \hat A_a + [\hat A_a,\hat A_b]
\qquad\textrm{and}\qquad
\hat D_a \hat\p \= \hat\pa_a\hat\p + [\hat A_a,\hat\p]\ .
\end{equation}
Like in the commutative case, $E_\th$ is minimized for a given $Q{\ge}0$
by the noncommutative BPS monopole configuration, which satisfies the
noncommutative Bogomolny equation
\begin{equation}
\hat F_{ab} \= \eps_{abc} \hat D_c \hat\p \ .
\end{equation}
Our main task is a systematic construction of solutions to this equation
by way of a Riemann-Hilbert problem.
 
\bigskip

\section{Monopoles via Riemann-Hilbert problems}

\subsection{Commutative case}

\noindent
{\bf Reduction of SDYM.} 
The Bogomolny equations (\ref{bog.eq.}) in $\R^3$ can be obtained by
dimensional reduction from the self-dual Yang-Mills (SDYM) equations in 
the Euclidean space $\R^4$,
\begin{equation}\label{sdym0}
*F\ =\ F 
\end{equation}
where $*$ denotes the Hodge star operator. 
Concretely, one puts $A_4=:\p$ and demands that all fields be translational
invariant along the $x^4$-axis,
\begin{equation}
T\,(A_a,\p)\ :=\ \pa_4 (A_a,\p)\ =\ 0 \qquad\textrm{for}\quad a=1,2,3 \ .
\end{equation}
This observation helps to show the integrability of the Bogomolny equations.
More explicitly, powerful methods for solving the SDYM equations can be applied
towards constructing monopole solutions as well.
In this section, we briefly describe one of these techniques. It reduces 
the task to a parametric Riemann-Hilbert problem, whose solutions for
various situations will be presented in subsequent sections.
{}From a higher mathematical vantage point, this framework has a geometric
description in terms of twistor and minitwistor spaces which is discussed 
in appendices A, B and~C.

\noindent
{\bf Linear system.}
The SDYM equations can be reformulated as the compatibility condition
of two linear differential equations for an auxiliary $n\times n$ matrix 
function $\psi(x,\la )$~\cite{Ward3, BZ}. The latter depends 
holomorphically on a new variable (`spectral parameter') $\la$ which lies 
in the extended complex plane $\C P^1=\C\cup\{\infty\}$. 
Upon dimensional reduction from $\R^4$ to $\R^3$ of this linear system of 
equations one obtains a reduced linear system whose compatibility condition
yields the Bogomolny equations. It is, however, advantageous not to discard
the ignorable coordinate $x^4$ prematurely but to allow $\psi(x,\la )$
to still depend on it.

To become concrete, let us introduce complex coordinates
\begin{equation} \label{Coco}
y=x^1+\im x^2\  ,\quad z=x^3-\im x^4\ ,\quad
\bar y=x^1-\im x^2\  ,\quad \bar z=x^3+\im x^4
\end{equation}
and put
\begin{equation} 
A_y=\sfrac{1}{2}(A_1-\im A_2)\ ,\quad
A_z=\sfrac{1}{2}(A_3+\im \p)\ ,\quad
A_{\bar y}=\sfrac{1}{2}(A_1+\im A_2)\ ,\quad
A_{\bar z}=\sfrac{1}{2}(A_3-\im \p)\ .
\end{equation}
With $D_a:=\pa_a+A_a$ and $D_4:=\pa_4+\p$ and furthermore $(A_a,\p )$ being
independent of $x^4$, the Bogomolny equations (\ref{bog.eq.}) read
\begin{equation}\label{Sdym2}
[D_y, D_z]=0\quad ,\qquad [D_{\bar y},D_{\bar z}]=0\quad ,\qquad
[D_y, D_{\bar y}]+[D_z, D_{\bar z}]=0\ .
\end{equation}
As advertized, they can be obtained as the compatibility condition of the 
linear system
\begin{equation}\label{linsys}
(D_{\bar y} - \la D_z )\,\psi (x,\la )\=0 \qquad\mbox{and}\qquad
(D_{\bar z} + \la D_y )\,\psi (x,\la )\=0\ , 
\end{equation}
where the argument $x$ in the auxiliary $n\times n$ matrix function 
$\psi (x,\la)$ still stands for $(x^1,x^2,x^3,x^4)$ or 
$(y,\bar y,z,\bar z)$.

\noindent
{\bf Riemann-Hilbert problems.}
The extended complex plane $\C P^1$ can be covered by two coordinate patches 
$U_+$ and $U_-$ with
\begin{equation}\label{cpuu}
\C P^1=U_+\cup U_- \quad ,\qquad
U_+=\C P^1\setminus\{\infty\} \quad ,\qquad
U_-=\C P^1\setminus\{0\} \ ,
\end{equation}
and coordinates $\la$ and $\lt$ on $U_+$ and $U_-$, respectively.
On the intersection $U_+\cap U_-\simeq\C^*$ these coordinates are
related by
\begin{equation}\label{llt}
\la \=\lt^{-1}\ .
\end{equation}

Suppose we are given an $n\times n$ matrix $f_{+-} (x^1,x^2,x^3,\la)$ 
which is a regular real-analytic function of
\begin{equation}
\la\ \in\ S^1\ \subset\ U_+\cap U_-\ ,
\end{equation}
i.e. it is defined on a circle for any fixed $(x^1,x^2,x^3)$ from an open set 
in $\R^3$. Then we may consider a parametric Riemann-Hilbert problem:
for each fixed $(x^1,x^2,x^3)$ try to factorize this matrix-valued function,
\begin{equation}\label{fp+p-}
f_{+-}(x^1,x^2,x^3,\la)\ =\ \psi_+^{-1}(x,\la)\,\psi_-(x,\la) \ ,
\end{equation}
in such a way that the two matrix factors $\psi_+$ and $\psi_-$ on the right
hand side are boundary values of holomorphic functions on (subsets of)
$U_+$ and $U_-$, respectively. 
Note that $\psi_{\pm}$ in general depend also on~$x^4$ while $f_{+-}$ does not.

Suppose further that the matrix-valued function $f_{+-}$ is chosen to depend 
on $(x^1,x^2,x^3)$ only through the combination
\begin{equation}\label{hhll}
\h \= y-2\la x^3 -\la^2{\bar y} 
\= (1{-}\la^2)x^1 +\im(1{+}\la^2)x^2 -2\la x^3 \qquad\mbox{for}\quad
\la \in {S^1} \subset U_+ \cap   U_- 
\end{equation}
in a holomorphic fashion, i.e. we have $f_{+-}(\h,\la)$ only.\footnote{
Our notation is slightly sloppy: We denote both $f_{+-}(\eta,\la)$
and $f'_{+-}(x,\la)\equiv f_{+-}(\eta(x,\la),\la)$ by the same symbol.}
{}From $\pa_{\bar\h}f_{+-}=0=\pa_{\lb}f_{+-}$ 
and the expression (\ref{hhll}) for $\h$  it follows that
\begin{equation}\label{holcond}
(\pa_{\bar y}-\la\pa_z)f_{+-}\=(\pa_{\bar z}+\la\pa_y)f_{+-}\=0\ .
\end{equation}
Substituting (\ref{fp+p-}) into (\ref{holcond}) and using the (generalized)
Liouville theorem, we obtain
$$
\psi_{+}(\pa_{\bar y}-\la\pa_z)\psi_{+}^{-1}\ =\
\psi_{-}(\pa_{\bar y}-\la\pa_z)\psi_-^{-1}\ =\
\sfrac{1}{2}(A_1 +\im A_2)-\sfrac{\la}{2}(A_3 +\im \p )\ ,
$$
\begin{equation}\label{AA1}
\psi_{+}(\pa_{\bar z}+\la\pa_y)\psi_{+}^{-1}\ =\
\psi_{-}(\pa_{\bar z}+\la\pa_y)\psi_-^{-1}\ =\
\sfrac{1}{2}(A_3 - \im \p ) + \sfrac{\la}{2}(A_1 - \im A_2)\ ,
\end{equation}
where the right hand sides are linear in~$\la$ and define four
functions $A_a(x)$ and~$\p(x)$ independent of~$x^4$.
Hence, starting with a parametric Riemann-Hilbert problem on 
$S^1\subset \C P^1$, we arrived to the linear system (\ref{linsys})
with a matrix function $\psi_+$ or $\psi_-$ instead of $\psi$.
By putting $\la =0$ we get 
\begin{equation} \label{AAAA1}
\sfrac{1}{2}(A_1+\im A_2)\ =\ 
\psi_+(\la)\pa_{\bar y}\psi_+^{-1}(\la)|_{\la=0}\ ,\qquad
\sfrac{1}{2}(A_3-\im\p)\ =\ 
\psi_+(\la)\pa_{\bar z}\psi_+^{-1}(\la)|_{\la=0}\ .
\end{equation}
Similarly, we may take the limit $\la{\to}\infty$ to obtain
\begin{equation}\label{AAAA2}
\sfrac{1}{2}(A_1 - \im A_2)\ =\ 
\psi_-({\la})\pa_y\psi_-^{-1}({\la})|_{\la =\infty}\ ,\qquad
\sfrac{1}{2}(A_3 +\im \p )\ =\ 
\psi_-({\la})\pa_z\psi_-^{-1}({\la})|_{\la =\infty}\ .
\end{equation}
In this way, splitting the matrix-valued function $f_{+-}$ automatically
yields solutions $(A_a,\p)$ to the Bogomolnyi equations~(\ref{bog.eq.}). 

\noindent
{\bf Reality conditions.}
The components of a gauge potential and the Higgs field constructed in 
this way will, however, in general not be antihermitean. In order to 
insure that $\ A_a^{\dagger}=-A_a\ $ and $\ \p^{\dagger}=-\p\ $ 
we may impose the following 
`reality' conditions on the matrices $f_{+-}$ and $\psi_{\pm}$:
\begin{equation}\label{Cond2}
f_{+-}^\+ (x, -{\bar\la}^{-1})\ =\ f_{+-} (x,\la)\ ,
\end{equation}
\begin{equation}\label{Cond1}
\psi_+^\+(x, - {\bar\la}^{-1})\ =\ \psi_-^{-1}(x, {\la})\ ,
\end{equation}
which are compatible with the Riemann-Hilbert problem.
Actually, the `reality' condition (\ref{Cond2}) on $f_{+-}$ 
already guarantees the possibility of~(\ref{Cond1}). Namely,
if $\psi_\pm$ do not satisfy (\ref{Cond1}) then
one may perform a {\it nonunitary\/} gauge transformation
\begin{equation}
\psi_{\pm} \mapsto \psi_{\pm}^g=g^{-1}\psi_{\pm}
\quad\textrm{inducing}\quad
A_a \mapsto A_a^g=g^{-1}A_a\, g +g^{-1}\pa_a\, g
\quad\mbox{and}\quad \p\mapsto\p^g=g^{-1}\p g\ ,
\end{equation}
where $g$ is to be computed from
\begin{equation}
g^2\ =\ \psi_-({\la})\,\psi_+^\+(-\bar\la^{-1})\ =\ 
\psi_+(\la)\,\psi_-^\+(-\bar\la^{-1})\ =\ (g^{\dagger})^2 
\end{equation}
and is independent of~$\la$. This gauge transformation leaves
$f_{+-}$ inert but produces `real' auxiliary functions $\psi_{\pm}^g$
and, hence, an antihermitean configuration $(A_a^g,\p^g)$.
The residual gauge freedom is of the same form but with $g'\in\mbox{U}(n)$.
To summarize, the task of constructing real monopole solutions
of the Bogomolny equations can be reformulated as finding solutions to 
a (parametric) Riemann-Hilbert problem supplemented by the 
`reality' conditions (\ref{Cond2}), (\ref{Cond1}). For more detailed
discussion and references see the appendices.

\subsection{Noncommutative case}

\noindent
In order to noncommutatively deform the above reformulation of 
the Bogomolny equations in terms of a Riemann-Hilbert problem,
we replace the coordinates $x^1$ and $x^2$ by operators $\xh^1$ and $\xh^2$
or, in complex notation, $y$ and $\yb$ by $\yh$ and $\hat\yb$, respectively.
In keeping with the notation for the commutative case, we introduce
the combinations
\begin{equation}
{\hat\h}\ =\ \yh -2\la x^3-\la^2{\hat{\yb}}
        \ =\ (1{-}\la^2)\xh^1 + \im(1{+}\la^2)\xh^2 - 2\la x^3
\end{equation}
and consider a matrix-valued function $\fh_{+-}({\hat\h},\la)$ 
depending analytically on $\la\in S^1=\{\la\in\C P^1: |\la |=1\}$.

Since all equations of the previous subsection are of matrix character
their generalization to the noncommutative situation is formally trivial:
all matrix entries are to be replaced by the corresponding operator-valued
entities. In other words, we simply substitute operators $\xh^a$ instead 
of coordinates $x^a$, retaining $x^3$ (and of course $\la$) as commutative.
Finally, a correspondence between solutions 
$\bigl(\hat{A}_a(\xh),\hat{\p}(\xh)\bigr)$ 
of the noncommutative Bogomolny equations and solutions 
$\bigl(\hat\psi_+(\xh,\la),\hat\psi_-(\xh,{\la})\bigr)$ 
of an operator Riemann-Hilbert problem for $\fh_{+-}({\hat\h},{\la})$ 
is described by formulae generalizing (\ref{AA1}), 
$$
\hat\psi_+(\pa_{\bar y}-\la\pa_z)\hat\psi_+^{-1}\ =\
\hat\psi_-(\pa_{\bar y}-\la\pa_z)\hat\psi_-^{-1}\ =\
\sfrac{1}{2}(\hat A_1 +\im \hat A_2)-\sfrac{\la}{2}(\hat A_3 +\im \hat\p)\ ,
$$
\begin{equation}\label{HatAA}
\hat\psi_+(\pa_{\bar z}+\la\pa_y)\hat\psi_+^{-1}\ =\
\hat\psi_-(\pa_{\bar z}+\la\pa_y)\hat\psi_-^{-1}\ =\
\sfrac{1}{2}(\hat A_3 - \im \hat\p) + \sfrac{\la}{2}(\hat A_1 - \im\hat A_2)\ .
\end{equation}
So, for a given splitting of an operator-valued function,
$ f_{+-}(\hat\h(\xh,\la),\la)=\hat\psi_+^{-1}(\xh,\la)\,\hat\psi_-(\xh,{\la})$,
one can calculate $(\hat A_a,\hat\p)$ from 
$$
\sfrac{1}{2}(\hat A_1 +\im \hat A_2)\ =\ 
\hat\psi_+(\la)\pa_{\bar y}\hat\psi_+^{-1}(\la)|_{\la=0}\quad,\qquad
\sfrac{1}{2}(\hat A_3 -\im \hat\p ) \ =\ 
\hat\psi_+(\la)\pa_{\bar z}\hat\psi_+^{-1}(\la)|_{\la=0}\quad,\phantom{XXXx}
$$
\begin{equation}\label{Opflds}
\sfrac{1}{2}(\hat A_1 -\im\hat A_2)\ =\
\hat\psi_-({\la})\pa_y\hat\psi_-^{-1}({\la})|_{\la=\infty}\ ,
\qquad
\sfrac{1}{2}(\hat A_3 +\im\hat\p ) \ =\
\hat\psi_-({\la})\pa_z\hat\psi_-^{-1}({\la})|_{\la=\infty}\ .
\end{equation}
By construction these fields satisfy the noncommutative Bogomolny equations.
Fore more details see the appendices. 

\bigskip

\section{Dirac monopoles} 
 
\subsection{Dirac monopoles on commutative $\R^3$} 
 
\noindent
Here we want to rederive the famous Dirac monopole solution~\cite{Dirac} 
via splitting of an appropriately chosen function 
$f_{+-}$ depending holomorphically on the coordinates $\h$ and~$\la$. 
 
\noindent 
{\bf Family of curves and circles.} 
Recall that the Dirac monopole is a solution defined not 
on the whole space $\R^3$ but on $\R^3\backslash\{0\}$. 
For a given point $(x^1,x^2,x^3)\in\R^3$ with $\{x^a\}\ne\{0\}$, 
$\h (x,\la)$ vanishes in two points $\la_1(x)$ and $\la_2(x)=-1/\lb_1(x)$ 
on $\C P^1$. Namely, from $\h (x,\la)=0$ we have 
\begin{equation}\label{ylx} 
y-2\la x^3 -\la^2\yb \= -(\la-\la_1) (\la-\la_2)\yb \=0 \qquad\mbox{at}\quad 
\la=\la_1\quad\mbox{and}\quad\la=\la_2\ , 
\end{equation} 
where  
\begin{equation}\label{l1l2} 
\la_1=-\frac{r+x^3}{\yb}\qquad\mbox{and}\qquad
\la_2= \frac{r-x^3}{\yb}=- \frac{1}{\lb_1} \qquad\mbox{with}\quad
r^2:=y\yb+x^3x^3\ .
\end{equation} 
Here and in the following we suppress the implicit $x$~dependence
of $\la_{1,2}$ for brevity.
We assume that $|\la_1|\le|\la_2|$ and exchange their labels if this is not so. 
Now, consider a circle $S^1$ in $\C P^1$ defined as 
\begin{equation}\label{S1} 
S^1\ =\  
\begin{cases} 
\{\ \la\in \C P^1 : & |\la| =1 \quad\mbox{if}\quad |\la_1|<1<|\la_2| \ \} \\ 
\{\ \la\in \C P^1 : & |\la-\ve\la_1|=1\ \mbox{with}\ 0{<}\ve{<}1 
                     \quad\mbox{if}\quad |\la_1|=|\la_2|=1 \ \}
\end{cases} 
\end{equation} 
where in the second case we have 
$|\la_1{-}\ve\la_1| < |\la{-}\ve\la_1| < |\la_2{-}\ve\la_1|$
since $|\la_1{-}\ve\la_1|=1{-}\ve\ $, $\ |\la{-}\ve\la_1|=1$ 
and $|\la_2{-}\ve\la_1|=1{+}\ve$.
So, the circle $S^1$ separates the points $\la{=}0$ and $\la{=}\la_1$
from the points $\la{=}\infty$ and $\la{=}\la_2$, i.e. the two zeros 
of the coordinate function $\h(x,\la)$ lie in different domains.
 
\noindent 
{\bf Function to be factorized.} 
In order to construct the Dirac monopole, 
we choose for our holomorphic function the simplest one possible,
\begin{equation}\label{trfun1} 
f_{+-}\=\frac{\la}{\h} \qquad \Longleftrightarrow \qquad 
f_{-+}\ :=\ (f_{+-})^{-1}\=\frac{\h}{\la} \ ,
\end{equation} 
defined for $\la\in U_+\cap U_-$ and $\h\neq0$.
Taking $f_{-+}$ to be linear in~$\h$ determines
the explicit $\la$ dependence by requiring the reality condition (\ref{Cond2}),
which in the variables $\h$ and~$\la$ takes the form
\begin{equation}\label{cond3} 
f^{\dagger}_{+-}(-\lb^{-2}\bar\h\,,-\lb^{-1})\=f_{+-}(\h,\la) \ .
\end{equation} 
 
\noindent 
{\bf Solutions of Riemann-Hilbert problems.} 
Let us consider $f_{-+}=\h/\la$ and restrict $\la$ to a circle $S^1$ as defined 
in (\ref{S1}) on which both $f_{+-}$ and $f_{-+}$ are well defined. 
The function $f_{-+}$ can be split on  $S^1$ as follows: 
\begin{equation} \label{split1} 
f_{-+}\=\frac{1}{\la}(y{-}2\la x^3{-}\la^2\yb)  
\=(\xi_+ +\la^{-1}y \xi_+^{-1})(\xi_+ -\la \xi_+^{-1}\yb)
\ =:\ (\psi_-^{S}(x,\la))^{-1}\,\psi_+^{S}(x,\la)\ , 
\end{equation} 
where  
\begin{equation}\label{xi+} 
\xi_+ \= (r-x^3)^{1/2} \= (\la_2\yb)^{1/2} \= (-\la_1^{-1}y)^{1/2} 
\end{equation}
with $r^2=x^a x^a=y\yb+x^3 x^3$.
In fact, eqs.~(\ref{split1}) give us an $x^4$ independent solution 
\begin{equation} \label{psipm1} 
\psi_+^{S} \= \xi_+ - \la \xi_+^{-1}\bar y 
\qquad\mbox{and}\qquad 
(\psi_-^{S})^{-1} \= \xi_+ + \la^{-1}y \xi_+^{-1}
\end{equation} 
of the Riemann-Hilbert problem on $S^1$. One can write down another  
solution 
\begin{equation} \label{split2} 
f_{-+}\=\frac{1}{\la}(y{-}2\la x^3{-}\la^2\yb)
\=(\yb\xi_- +\la^{-1} \xi_-^{-1})(\xi_- y - \la \xi_-^{-1}) 
\ =:\ (\psi_-^{N}(x,\la))^{-1}\,\psi_+^{N}(x,\la)\ , 
\end{equation} 
where 
\begin{equation}\label{psipm2} 
\psi_+^{N} \= \xi_- y - \la \xi_-^{-1} 
\qquad\mbox{and}\qquad 
(\psi_-^{N})^{-1} \= \yb\xi_- + \la^{-1}\xi_-^{-1}  
\end{equation} 
with
\begin{equation}\label{xi-} 
\xi_- \= (r+x^3)^{-1/2} \= (-\la_1\yb)^{-1/2} \= (\la_2^{-1}y)^{-1/2} \ . 
\end{equation} 
It is not difficult to see that both solutions $(\psi_+^{S}, \psi_-^{S})$ 
and $(\psi_+^{N}, \psi_-^{N})$ of the Riemann-Hilbert problem satisfy the 
reality condition~(\ref{Cond1}) and therefore lead to real 
(antihermitean) solutions $(A_a,\p )$ of the Bogomolny equations. 
 
\noindent 
{\bf Dirac monopole.} 
Substituting $\psi_+^{S}(x,\la{=}0)=\xi_+=(r{-}x^3)^{1/2}$ 
into eqs.~(\ref{AAAA1}), we obtain 
\begin{equation}\label{Ayb+} 
A_{\yb}^{S}\equiv\sfrac{1}{2}(A_1^{S}+\im A_2^{S})\= 
\xi_+\pa_{\yb}\xi_+^{-1} \= -\frac{y}{4r(r{-}x^3)}\ ,\quad 
A_3^{S}=0 \quad\mbox{and}\quad
\p^{S}\=\frac{\im}{2r}\ . 
\end{equation} 
{}From (\ref{Ayb+}) it follows that 
\begin{equation}\label{A+} 
A_1^{S}\=-\frac{\im x^2}{2r(r{-}x^3)}\quad,\qquad 
A_2^{S}\= \frac{\im x^1}{2r(r{-}x^3)} \qquad\mbox{and}\qquad 
A_3^{S}=0\ , 
\end{equation} 
which coincide with the components of the gauge potential of the 
Dirac monopole defined on 
\begin{equation}\label{RS}
\R^3_S\ :=\ \R^3\backslash\{x^1=x^2=0,\ x^3\ge 0\}
\end{equation} 
and with the singularity along the $x^3{>}0$ axis. Analogously,  
substituting $\psi_+^{N}(x,\la{=}0)=\xi_-y$ into eqs.(\ref{AAAA1}), 
we obtain 
\begin{equation}\label{Ayb-} 
A_{\yb}^{N}\equiv\sfrac{1}{2}(A_1^{N}+\im A_2^{N})\= 
\xi_-\pa_{\yb}\xi_-^{-1} \= \frac{y}{4r(r{+}x^3)}\ ,\quad 
A_3^{N}=0 \quad\mbox{and}\quad 
\p^{N}\=\frac{\im}{2r} 
\end{equation} 
and therefore 
\begin{equation}\label{A-} 
A_1^{N}\= \frac{\im x^2}{2r(r{+}x^3)}\quad,\qquad 
A_2^{N}\=-\frac{\im x^1}{2r(r{+}x^3)} \qquad\mbox{and}\qquad 
A_3^{N}=0\ , 
\end{equation} 
which coincide with the components of the Dirac monopole gauge 
potential well defined on 
\begin{equation}\label{RN}
\R^3_N\ :=\ \R^3\backslash\{x^1=x^2=0,\ x^3\le 0 \}\ , 
\end{equation}
i.e. everywhere on $\R^3\backslash\{0\}$ besides the negative $x^3$-axis. 
Note that $\R^3_S$ and $\R^3_N$ in (\ref{RS}) and (\ref{RN}) denote
the southern and northern patches, respectively, of a two-sphere $S^2$ 
surrounding the origin.
On the intersection of the above domains the configurations  
$(A^{N},\p^{N})$ and $(A^{S},\p^{S})$ are well defined 
and related by a transformation 
\begin{equation}\label{A+A-} 
A^{N}\=A^{S}+\diff\,\ln \bigl(\frac{\yb}{y}\bigr)^{1/2}
\qquad\textrm{and}\qquad \p^{N}\=\p^{S} 
\end{equation} 
which provides a global description of the Dirac monopole on 
$\R^3\backslash\{0\}$~\cite{Wu1}. 
 
\noindent 
{\bf Dirac multi-monopoles.} 
It is obvious that for obtaining abelian  multi-monopoles 
via solving Riemann-Hilbert problems, as a function to be factorized
one may take the product 
\begin{equation}\label{trfun2} 
f_{-+}^{(n)} \= \prod_{k=1}^n \frac{\h_k}{\la} \qquad\textrm{where}\quad 
\h_k \= (1{-}\la^2)(x^1{-}a^1_k)+\im(1{-}\la^2)(x^2{-}a^2_k)-2\la(x^3{-}a^3_k)
\ =:\ \h-h_k
\end{equation} 
and the $3n$ real parameters $(a^1_k,a^2_k,a^3_k)=(\va_k)$ define the positions
of $n$ monopoles in $\R^3$. We restrict $f_{-+}^{(n)}$ to a contour on $\C P^1$
which avoids all zeros $\la_{1,k}$ and $\la_{2,k}$ of~$\h_k$ so that 
$f_{-+}^{(n)}$ is invertible on it. Using formulae (\ref{split1})-(\ref{xi-})
and abbreviating the relative coordinates $x^b-a^b_k=:x^b_k$, 
we easily split the factors in (\ref{trfun2}) individually as~\footnote{
Note that in this section $x_k$ stands for $(x_k^a)$ and $x$ denotes $(x^a)$
since the splitting is independent of~$x^4$.}
\begin{equation}
\frac{\h_k}{\la} \= (\psi_{-}^{S}(x_k,\la))^{-1}\psi_{+}^{S}(x_k,\la)
                 \= (\psi_{-}^{N}(x_k,\la))^{-1}\psi_{+}^{N}(x_k,\la)
\end{equation}
and obtain
\begin{align}\label{trfun3}
f_{-+}^{(n)} &\= \biggl[ \prod_{k=1}^n \psi_{-}^{S}(x_k,\la) \biggr]^{-1}
                 \biggl[ \prod_{\ell=1}^n \psi_{+}^{S}(x_{\ell},\la) \biggr]
\ =:\ (\psi_{-}^{S,n}(x,\la))^{-1} \psi_{+}^{S,n}(x,\la) \nonumber\\[8pt]
             &\= \biggl[ \prod_{k=1}^n \psi_{-}^{N}(x_k,\la) \biggr]^{-1}
                 \biggl[ \prod_{\ell=1}^n \psi_{+}^{N}(x_{\ell},\la) \biggr]
\ =:\ (\psi_{-}^{N,n}(x,\la))^{-1} \psi_{+}^{N,n}(x,\la) \ .
\end{align}
Therefore, 
\begin{equation}\label{psi++} 
\psi_{+}^{S,n}(x,\la{=}0) \= \prod\limits_{k=1}^{n}\xi_+(x_k^a)
\ =:\ \xi_+^{(n)}  \qquad\textrm{and}\qquad
\psi_{+}^{N,n}(x,\la{=}0) \= \prod\limits_{k=1}^{n}\xi_-(x_k^a)y_k
\ =:\ \xi_-^{(n)}y
\end{equation} 
and analogously for $\psi_{-}^{S,n}$ and $\psi_{-}^{N,n}$. 
 
Substituting (\ref{psi++}) into (\ref{AAAA1}), we obtain 
$$ 
A_{\yb}^{S,n}\=\xi_{+}^{(n)}\pa_{\yb}(\xi_{+}^{(n)})^{-1}
\= \sum\limits_{k=1}^n \xi_{+}(x_k^a)\pa_{\yb}\xi_{+}^{-1}(x_k^a)
\ =:\ \sum\limits_{k=1}^n A_{\yb}^{S}(x_k^a)\ , 
$$ 
\begin{equation}\label{Ap} 
A_{3}^{S,n}\=0 \qquad\textrm{and}\qquad 
\p^{S,n}\=\sum\limits_{k=1}^n\frac{\im}{2r_k} \qquad\textrm{with}\quad
r_k^2 = \sum_{a=1}^3 x_k^a x_k^a \ ,
\end{equation} 
and analogously for $A_{\yb}^{N,n}$ with $A_{3}^{N,n}=0$ and  
$\p^{N,n}=\p^{S,n}$. 
Thus, solving Riemann-Hilbert problems on an appropriate circle 
surrounding all $\la{=}\la_{1,k}$, 
we obtain a linear superposition of $n$~Dirac monopole solutions 
regular on the space $\R^3\backslash\{\va_1,\ldots,\va_n\}$.

\subsection{Noncommutative Dirac monopoles} 
 
\noindent
Abelian solutions of the noncommutative Bogomolny equations have been 
constructed in closed form by Gross and Nekrasov~\cite{Gross:2000wc}
in the noncomutative generalization of the Nahm approach~\cite{Nahm1}.  
Here we show how their solution (the noncommutative Dirac monopole) can be 
obtained by solving the appropriate Riemann-Hilbert problem. Note that the 
string-like singularity of the commutative Dirac monopole is absent 
in the noncommutative configuration~\cite{Gross:2000wc}. 
For this reason we can and will restrict ourselves to a  
noncommutative generalization of the southern patch quantities
$\psi_{\pm}^{S}$, $\xi_+$ etc.~derived in section 4.1 and in the following 
omit the `$S$' superscript indicating the singularity of the standard Dirac 
monopole on the positive $x^3$-axis. 
 
\noindent 
{\bf Proper circles.} 
In order to pass to the noncommutative Riemann-Hilbert problem we promote
the coordinates $(y,\yb)$ to operators $(\yh,\hat{\yb})$ acting in a 
one-oscillator Fock space~$\Hcal$ from the left or in its dual~$\Hcal^*$
from the right. To mimic the commutative construction we should
first of all describe circles $S^1\subset\C P^1$ such that the operator
\begin{equation}\label{hath} 
\hat{\h} \= \yh - 2\la x^3 - \la^2\hat{\yb} 
\end{equation} 
does not vanish on vectors from $\Hcal$ or $\Hcal^*$ if $\la\in S^1$. 
Surprisingly, in the noncommutative case the situation is simpler than  
the one discussed in section 4.1. Namely, recall that $\yh$ is proportional 
to the annihilation operator $a$ and that $\hat{\yb}$ is proportional  
to the creation operator $a^\+$. Obviously, the operator $\hat{\h}$ is 
a linear combination of $a$, $a^\+$, and ${\bf 1}$. 
Therefore, its eigenvectors $|\vp\>\in\Hcal$, 
\begin{equation}\label{hhp} 
\hat{\h}|\vp\> \= \vp |\vp\> \qquad\textrm{with}\quad \vp\in\C\ , 
\end{equation} 
are so-called squeezed (or generalized coherent) 
states~\cite{Perelomov, Blaizot}. It is well known that its zero eigenvalues,
$\vp=0$, belong to normalizable states (and therefore occur in~$\Hcal$)
if and only if $|\la|<1$~\cite{Perelomov, Blaizot}. 
Analogously, the kernel of $\hat{\h}$ in $\Hcal^*$, $\<\vp'|\hat\h =0$,
is non-empty iff $|\la|>1$.
Therefore, on the circle 
\begin{equation}\label{ss1} 
S^1\ =\ \{\la\in\C P^1: |\la |=1\} 
\end{equation} 
the operator $\hat{\h}$ has no zero modes either on $\Hcal$ or on $\Hcal^*$. 
So, for $|\la|=1$ the operator $\hat\h$ is invertible 
on states with finite norm, i.e. on $\Hcal$ as well as on $\Hcal^*$.
  
\noindent 
{\bf Noncommutative Dirac monopole.} 
We adhere to the simplest ansatz and assume that the holomorphic function 
(\ref{trfun1}) describing the commutative Dirac monopole 
keeps its form in the noncommutative case, i.e. 
\begin{equation}\label{trfun4} 
\fh_{+-}\= \la\,\hat{\h}^{-1} \qquad \mbox{and} \qquad 
\fh_{-+}\= \la^{-1}\,\hat{\h}\ . 
\end{equation} 
In (\ref{trfun4}) we assume $|\la|=1$ so that $\hat{\h}$ is invertible. 
Again, the operator (\ref{trfun4}) is real in the sense that 
\begin{equation}\label{cond4} 
\fh_{+-}^\+(-\hat{\h}^\+/\lb^2 , -1/{\lb}) \= \fh_{+-}(\hat{\h},{\la}) 
\end{equation} 
(cf.~eq.~(\ref{cond3})). 
 
The operator-valued function 
\begin{equation}\label{trfun5} 
\fh_{-+}\= \frac{\hat{\h}}{\la}\=\frac{1}{\la}(\yh-2\la x^3-\la^2\hat{\yb})
\= \sqrt{2\th}\,\bigl(\la^{-1}a - \frac{2x^3}{\sqrt{2\th}}-\la a^\+ \bigr) 
\end{equation} 
can be split as follows:\footnote{
We are grateful to N. Nekrasov for proposing this splitting.} 
\begin{equation}\label{split3} 
{\fh}_{-+}\=(\xi +\la^{-1}\hat{y} \xi^{-1})\,(\xi - \la \xi^{-1}\hat{\yb})
\ =:\ \ph_-^{-1}(\xh,\la)\,\ph_+(\xh,\la)\ , 
\end{equation} 
where the operator $\xi$ is implicitly defined by the equation 
\begin{equation}\label{xi} 
\xi^2 - \yh\xi^{-2}\hat{\yb} \= -2x^3 \ . 
\end{equation} 
This equation has been solved earlier in~\cite{Gross:2000wc}. Namely,  
\begin{equation}\label{solGN} 
\xi(x^3)\=(2{\th})^{1/4}\sum_{\ell\ge0} \xi_\ell(x^3)\,|\ell\>\<\ell| 
\qquad\mbox{with}\qquad  
\xi_\ell(x^3) = \sqrt{ \frac{\Ups_{\ell-1} (\frac{x^3}{\sqrt{2\th}})} 
                            {\Ups_{\ell}   (\frac{x^3}{\sqrt{2\th}})} } \ , 
\end{equation} 
where the special functions $\Ups_{\ell}(x)$ are given by formulae 
\begin{equation}\label{Ups} 
\Ups_{-1}=1 \qquad\textrm{and}\qquad 
\Ups_\ell(x)=\int_{0}^{\infty}\!\diff p\  
\frac{p^\ell}{\ell !}\,\exp{\bigl(2xp-\sfrac{1}{2}p^2\bigr)}
\qquad\mbox{for}\quad \ell\ge 0\ . 
\end{equation} 
One can show that $\Ups_\ell (x)$ satisfy the recurrence relations 
\begin{equation}\label{recrel} 
(\ell{+}1)\Ups_{\ell+1}\=2x\Ups_{\ell} + \Ups_{\ell-1} 
\end{equation} 
and also 
\begin{equation}\label{rel} 
\pa_x\Ups_{\ell}(x)\=2(\ell{+}1)\Ups_{\ell+1}(x)\ . 
\end{equation} 
One easily sees that (\ref{solGN}) solves (\ref{xi}) due to  
eqs.~(\ref{recrel})~\cite{Gross:2000wc}.  
 
{}From (\ref{split3}) we obtain operators 
\begin{equation}\label{ops} 
\ph_+(\xh,\la) \= \xi - \la\xi^{-1}\hat{\yb} 
\qquad\mbox{and}\qquad 
\ph_-^{-1}(\xh,\la) \= \xi + \la^{-1}\hat{y}\xi^{-1} 
\end{equation} 
satisfying the reality condition (\ref{Cond1}). 
Substituting (\ref{ops}) into formulae (\ref{Opflds}), we obtain  
a solution $(\hat{A}_a,\hat{\p})$ of the noncommutative Bogomolny 
equations: 
\begin{equation}\label{opsA} 
\hat{A}_1\= -\frac{1}{\th}\,[\xh_1,\xi]\,\xi^{-1}\quad,\qquad  
\hat{A}_2\=  \frac{1}{\th}\,[\xh_2,\xi]\,\xi^{-1}\quad,\qquad  
\hat{A}_3\=0 \ ,
\end{equation} 
\begin{equation}\label{opp} 
\hat{\p}\= -\im (\pa_3\xi)\,\xi^{-1}
\= -\im\sum_{\ell\ge0}\pa_3 (\ln\xi_\ell(x^3))\,|\ell\>\<\ell| \ . 
\end{equation} 
This describes the noncommutative Dirac monopole. 
For a discussion of its properties and its $D$-brane interpretation 
see~\cite{Gross:2000wc, Nekrasov:2000ih}. 
 
\noindent 
{\bf  Noncommutative Dirac multi-monopoles.} 
In order to obtain noncommutative multi-mono\-poles 
one may consider the Riemann-Hilbert problem for the operator-valued function 
\begin{equation}\label{f-+n} 
\fh_{-+}^{(n)}\=\prod_{k=1}^n \frac{1}{\la}\,(\hat\h-h_k)\ , 
\end{equation} 
where $h_k$ are polynomials in $\la$ not depending on the coordinates~$x$ 
and satisfying 
\begin{equation}\label{gamma} 
\overline{h_k(-1/\bar\la)}\=-\la^{-2}\,h_k(\la)
\end{equation} 
with the bar denoting complex conjugation. The simplest example of such 
$h_k$ is written down in eq.~(\ref{trfun2}). 
 
Note that it is not easy to split the operator-valued function (\ref{f-+n})
even in the simplest $n{=}2$ case. We will not try to do this here.
Instead, in the next section we describe noncommutative generalizations
of SU(2) Wu-Yang multi-monopole solutions which asymptotically 
(for large separation) approach to Dirac multi-monopole solutions 
with the U(1) group embedded into the gauge group SU(2). The exact equivalence
of the Dirac multi-monopoles to the general SU(2) Wu-Yang multi-monopoles
generated by the ${\cal A}_n$ Atiyah-Ward ansatz is proven in appendix~C.
 
\newpage

\section{Wu-Yang monopoles}

\subsection{The Wu-Yang SU(2) monopole}

\noindent
A spherically-symmetric singular monopole solution
of the SU(2) gauge field equations has been obtained by Wu and Yang in 
1969~\cite{Wu2}. It was later interpreted (see e.g.~\cite{Actor} and 
references therein) as a static solution,
\begin{equation}\label{Aphi}
A_a\= \eps_{abc}\,\frac{\s_c}{2\im}\,\frac{x^b}{r^2}
\quad ,\qquad
\p \= \frac{\s_a}{2\im}\,\frac{x^a}{r^2} \ ,
\end{equation}
of the Yang-Mills-Higgs equations (\ref{em}). It is not difficult to see 
that this configuration satisfies also the Bogomolny 
equations (\ref{bog.eq.}) but has infinite energy. Initially it was 
thought that (\ref{Aphi}) is genuinely nonabelian, 
yet after 't~Hooft's analysis~\cite{'tHooft} it was realized that 
it is nothing but the (abelian) Dirac monopole in disguise 
(see e.g.~\cite{Arafune, Wu1, Bais, Corrigan1}).
Furthermore, the gauge potential of the finite-energy spherically symmetric 
BPS SU(2) monopole approaches the gauge potential in~(\ref{Aphi}) for large~$r$.

The equivalence of the Wu-Yang SU(2) monopole and the Dirac U(1)
monopole (with one gauge group embedded into the other)
can be seen as follows. Recall that the Dirac monopole
is described (see section 4) by the gauge potentials $A^{S}$
and $A^{N}$ defined on $\R^3_S$ and $\R^3_N$, respectively.
Here, $\R^3_S\cup \R^3_N=\R^3\backslash\{0\}$, and on the overlap region
$\R^3_S\cap \R^3_N$ the gauge potentials are related via a transition function
(cf.~(\ref{A+A-}))
\begin{equation}\label{fD}
f_{NS}^{(D)}=\Bigl(\frac{y}{\yb}\Bigr)^{1/2}=: \e^{i\vp}
\end{equation}
as
\begin{equation}\label{AN}
A^{N}\=f_{NS}^{(D)}A^{S}(f_{NS}^{(D)})^{-1} + 
f_{NS}^{(D)}\diff (f_{NS}^{(D)})^{-1}\ .
\end{equation}
Let us multiply this equation by the matrix
$\s_3=\left(\begin{smallmatrix}1&\ 0\\0&-1\end{smallmatrix}\right)$ 
and rewrite it as 
\begin{equation}\label{ANs3}
A^{N}\s_3\ =\ f_{NS}\ A^{S} \s_3\ (f_{NS})^{-1}\ +\ 
f_{NS}\ \diff (f_{NS})^{-1}\ ,
\end{equation}
where
\begin{equation}\label{fNS}
f_{NS}\ :=\
\begin{pmatrix}
f_{NS}^{(D)}&0\\0&(f_{NS}^{(D)})^{-1}
\end{pmatrix}
\=\begin{pmatrix}
(\frac{y}{\yb})^{1/2}&0\\0&(\frac{y}{\yb})^{-1/2}
\end{pmatrix} \ .
\end{equation}
Obviously, (\ref{ANs3}) is equivalent to (\ref{A+A-}).

Note now that one may split the transition matrix (\ref{fNS}) as
\begin{equation}\label{fgNgS}
f_{NS}\ =\ g_N^{-1}\,g_S\ ,
\end{equation}
where the $2{\times}2$ unitary matrices
\begin{equation}\label{gNgS}
g_{N} :=\frac{1}{\sqrt{2r(r{+}x^3)}}
\begin{pmatrix}
\yb&r{+}x^3\\-(r{+}x^3)&y
\end{pmatrix} 
\qquad\mbox{and}\qquad 
g_{S} :=\frac{1}{\sqrt{2r(r{-}x^3)}}
\begin{pmatrix}
r{-}x^3&\yb\\-y&r{-}x^3
\end{pmatrix}
\end{equation}
are well defined on $\R^3_N$ and $\R^3_S$, respectively.
Substituting (\ref{fgNgS}) into (\ref{ANs3}), we obtain
\begin{equation}\label{gAg}
g_{N}\, A^{N} \s_3\, g_{N}^{-1}\ +\ 
g_{N}\, \diff g_{N}^{-1}\ =\ 
g_{S}\, A^{S} \s_3\, g_{S}^{-1}\ +\ 
g_{S}\, \diff g_{S}^{-1}\ =:\ 
A^{SU(2)}\ ,
\end{equation}
where by construction $A^{SU(2)}$ is well defined on 
$\R^3_S\cup \R^3_N=\R^3\backslash\{0\}$. Geometrically, the existence
of the splitting (\ref{fgNgS}) means that  
the Dirac monopole's nontrivial U(1) bundle trivializes 
when being imbedded into an SU(2) bundle. The matrices (\ref{gNgS})
define this trivialization since $f_{NS}\ \mapsto\ f_{NS}^{\textrm{new}}=
g_N\, f_{NS} g_{S}^{-1}={\bf 1}_2$~\cite{Wu1}.

Substituting (\ref{gNgS}) and the Dirac monopole configuration 
(\ref{A+}) and (\ref{A-}) into (\ref{gAg}), we obtain
$A^{SU(2)}=A_{\yb}\diff{\yb} + A_y \diff y + A_3 \diff x^3$ with
\begin{equation}\label{AybAy}
A_{\yb} =\frac{1}{{4r^2}}
\begin{pmatrix}
-y&2x^3\\0&y
\end{pmatrix} 
\quad ,\qquad 
A_{y} =\frac{1}{{4r^2}}
\begin{pmatrix}
-{\yb}&0\\2x^3&\yb
\end{pmatrix} 
\qquad\mbox{and}\qquad 
A_{3} =\frac{1}{{2r^2}}
\begin{pmatrix}
0&-{\yb}\\y&0
\end{pmatrix} 
\end{equation}
which in real coordinates coincides with $(A_a)$ given in (\ref{Aphi}).
Analogously, for the Higgs field we have
\begin{equation}\label{pNpS}
\p^{N}\s_3\=g_{N}^{-1}g_{S}\,\p^{S}\s_3\, g_{S}^{-1}g_{N}
\quad\Longrightarrow\quad
\p\ :=\ g_{N}\, \p^{N}\s_3\, g_{N}^{-1} 
     \= g_{S}\, \p^{S}\s_3\, g_{S}^{-1}
\= \frac{x^a}{r^2}\frac{\s_a}{2\im}\ ,
\end{equation}
which is identical to $\p$ in (\ref{Aphi}). 

The equivalence of the Wu-Yang SU(2) monopole to the Dirac abelian monopole
extends to the multi-monopole situation. 
One may take, for instance, the abelian multi-monopole solution (\ref{Ap}) 
and transform it into a Wu-Yang multi-monopole solution. 
Explicit matrices $g_N$ and $g_S$ may be found in~\cite{Arafune, Bais} 
for the case of monopoles situated along a straight line.
Instead of directly generalizing the formulae presented above, we shall
alternatively derive special Wu-Yang multi-monopole configurations 
via solving a Riemann-Hilbert problem in the following subsection. 
A discussion of the general Wu-Yang SU(2) multi-monopole and its equivalence 
to the Dirac U(1) multi-monopole is relegated to appendix~C.

\subsection{Wu-Yang SU(2) multi-monopoles}

\noindent
{\bf Matrix-valued holomorphic function.}
We would like to formulate a matrix Riemann-Hilbert problem appropriate for
the Wu-Yang monopole. Here, the main skill consists in guessing a suitable
holomorphic matrix-valued function $f_{+-}^{SU(2)}$. 
We make the Atiyah-Ward ${\cal A}_1$ ansatz~\footnote{
For a discussion of the more general ${\cal A}_n$ Atiyah-Ward ansatz and
the corresponding Wu-Yang multi-monopole see appendix~C.} 
\cite{Atiyah2,Ward-Wells}
\begin{equation}\label{fSU2}
f_{+-}^{SU(2)}\=
\begin{pmatrix}
\r &\la^{-1}\\ -\la &0
\end{pmatrix}
\end{equation}
containing a function $\r=\r(\eta,\la)$ holomorphic in the (local)
coordinates $\eta$ and $\la$.
We restrict $f_{+-}^{SU(2)}$ to our circle $S^1\subset\C P^1$;
then $\r$ depends parametrically on $(x^1,x^2,x^3)\in U\subset\R^3$.
Furthermore, we impose on $\r$ the reality condition
\begin{equation}\label{reco}
\r^{\+}(x, -{\lb}^{-1})\=\r (x,\la)
\end{equation}
which guarantees that $f_{+-}^{SU(2)}$ satisfies the reality
condition (\ref{Cond2}) and therefore will produce
real solutions of the Bogomolny equations.

It is useful to expand $\r$ in a Laurent series in $\la$,
\begin{equation}\label{rsum}
\r\ = \sum\limits^\infty_{m=-\infty}\r_m\,\la^m\ =\ \r_-+\r_0 +\r_+\ ,
\end{equation}
where
\begin{equation}\label{rrsum}
\r_+:=\sum\limits_{m >0}\r_m\,\la^m \qquad\mbox{and}\qquad 
\r_-:=\sum\limits_{m <0}\r_m\,\la^m \ .
\end{equation}
The Laurent coefficients $\r_m=\r_m(x^1,x^2,x^3)$ 
are not functionally independent.
{}From the restricted $x$-dependence of $\r$ (its holomorphicity in~$\h$) 
it follows that (c.f.~(\ref{holcond}))
\begin{equation}\label{horho}
(2\pa_{\yb} - \la\pa_3)\r\ =\ (\pa_3 + 2\la\pa_y )\r \ =\ 0
\end{equation}
which implies the recursion relations
\begin{equation}\label{parr}
2\pa_{\yb}\r_{m+1}\ =\ \pa_3\r_m\ =\ -2\pa_y\r_{m-1}\ .
\end{equation}

\noindent
{\bf Riemann-Hilbert problem and solutions.}
It is not hard to see that $f_{+-}^{SU(2)}$ can be split as
\begin{equation}\label{fpm}
f_{+-}^{SU(2)}\ =\ \psi_+^{-1}\,\psi_-\ ,
\end{equation}
where (see also \cite{LP1})
\begin{equation}\label{-ps-}
\psi_-^{-1}\=
\begin{pmatrix} 1&-\la^{-1}\\-\la\r_-& \r_0 +\r_- \end{pmatrix}
\frac{1}{\sqrt{\r_0}}
\quad ,\qquad
\psi_-\=\frac{1}{\sqrt\r_0}
\begin{pmatrix}\r_0 +\rho_-&\la^{-1}\\ \la\rho_-  & 1\end{pmatrix}\ , 
\end{equation}
\begin{equation}\label{+ps+}
\psi_+\=\frac{1}{\sqrt\r_0}
\begin{pmatrix}1&\la^{-1}\rho_+\\ \la& \r_0 +\rho_+\end{pmatrix}
\quad , \qquad 
\psi_+^{-1}\=
\begin{pmatrix}\r_0 +\rho_+&-\la^{-1}\rho_+\\-\la& 1\end{pmatrix} 
\frac{1}{\sqrt\r_0}\ .
\end{equation}
Formulae (\ref{fpm}) -- (\ref{+ps+}) solve our parametric Riemann-Hilbert 
problem on the circle~$S^1$ in an $x^4$-independent manner.
One can show that the above matrices $\psi_{\pm}(x,\la)$ 
satisfy the reality condition (\ref{Cond1}).
Substituting (\ref{+ps+}) into formulae (\ref{AAAA1}) and using the recursion 
relations (\ref{parr}), we get
\begin{equation}\label{Aandp}
A_a\= -\eps_{abc}\,\frac{\s_c}{2\im}\,\r_0^{-1}\,\pa_b\r_0
\qquad\mbox{and}\qquad
\p \= -\frac{\s_a}{2\im}\,\r_0^{-1}\,\pa_a\r_0\ ,
\end{equation}
where, due to (\ref{parr}), the function $\r_0$ is a solution 
of the Laplace equation
\begin{equation}\label{pa2r}
(\pa_1^2 + \pa_2^2 + \pa_3^2)\r_0 \= 0
\end{equation}
on an open subset $U$ of $\R^3$. 

\noindent
{\bf Charge one monopole.} 
On $U=\R^3\backslash\{0\}$, the only spherically symmetric solution
(up to an additive constant and irrelevant normalization) is
\begin{equation}\label{r0}
\r_0=\frac{1}{2r} \ .
\end{equation}
Substitution it into (\ref{Aandp}) immediately yields
the Wu-Yang monopole solution (\ref{Aphi}).

It is noteworthy that one can find the explicit form not only
of $\r_0$ but also of the complete holomorphic function $\r$. Therefore,
we can give a closed expression for the matrix-valued function 
$f_{+-}^{SU(2)}$ in (\ref{fSU2}) which describes the Wu-Yang monopole. 
In order to reconstruct the function $\r$ from $\r_0$, 
we recall the reality condition~(\ref{reco}) as
\begin{equation}\label{reco2}
\r^{\+}(-{\lb}^{-2}\bar\h\, , -{\lb}^{-1}) \= \r(\h,\la) \ .
\end{equation}
Since the single-monopole solution~(\ref{Aphi}) contains no free (dimensionful)
parameters such as a scale or a position, this condition implies that $\r$ is 
a function of the combination $\frac{\h}{\la}=\la^{-1}y{-}2x^3{-}\la\yb$ only. 
The length dimension of $\r_0$ in~(\ref{r0}) then determines $\r$ to be 
proportional to~$(\frac{\h}{\la})^{-1}$. 

For explicit verification we fix the point $x\in\R^3\setminus\{0\}$ 
which yields $\la_{1,2}$ according to~(\ref{l1l2}).
Assuming as in section 4.1 that $|\la_1|<|\la|<|\la_2|$, we obtain
\begin{align}
\frac{\la}{\eta} 
&\= \frac{\la}{(\la-\la_1)(\la_2-\la)\,{\yb}} 
 \= \frac{1}{(\la_2{-}\la_1)\,{\yb}}\,
 \biggl(\frac{\la_1}{\la-\la_1}+\frac{\la_2}{\la_2-\la}\biggr) \nonumber\\[8pt]
&\= \frac{1}{(\la_2{-}\la_1)\,{\yb}}\,
 \biggl(\frac{\la_1/\la}{1-\la_1/\la}+\frac{1}{1-\la/\la_2}\biggr)
 \= \frac{1}{(\la_2{-}\la_1)\,{\yb}}\,\biggl\{ 1 +
 \sum_{\ell\ge 1}\Bigl(\frac{\la_1}{\la}\Bigr)^\ell +
 \sum_{\ell\ge 1}\Bigl(\frac{\la}{\la_2}\Bigr)^\ell\biggl\}
\end{align}
and therefore indeed
\begin{equation}\label{rl0} 
\Bigl(\frac{\la}{\eta}\Bigr)_0 \= \frac{1}{(\la_2{-}\la_1)\,{\yb}} 
\= \frac{1}{2r} \= \r_0 \ .
\end{equation}
Alternatively,
\begin{equation}
\r_0\=\oint\limits_{|\la|=1} \!\frac{\diff\la}{2\pi\im\la}\,\r
\={\textrm{res}}_{\la{=}\la_1} \h^{-1} \= \frac{1}{(\la_2{-}\la_1)\,\yb}
\= \frac{1}{2r} \ .
\end{equation}
Analogously for the other case in (\ref{S1}), $|\la_1|=|\la_2|=1$, 
one arrives at the equivalent solution $\r_0 = (2(1{+}\ve)r)^{-1}$
with $0<\ve<1$.
It is amusing that only the power~$-1$ of $\frac{\h}{\la}$ yields
a spherically symmetric function~$\r_0$.
We conclude that the Wu-Yang monopole is fully described by the function
\begin{equation}\label{rle}
\r(\h,\la) \= \frac{\la}{\eta} \= \bigl(\la^{-1}y-2x^3-\la\,\yb\bigr)^{-1} \ .
\end{equation}
The equivalence of the Wu-Yang solution to the Dirac monopole
solution follows also from the fact that this function 
coincides with the function (\ref{trfun1}) generating the Dirac monopole.

\noindent
{\bf Multi-monopole.}
To generate a multi-monopole solution describing $n$ monopoles
sitting at points $(a^1_k,a^2_k,a^3_k)=(\vec{a}_k)$ with $k=1,\ldots,n$, 
one may take the function
\begin{equation}\label{rn} 
\r^{(n)}\ =\ \sum_{k=1}^n\, \frac{\la}{\eta_k} \qquad\textrm{with}\quad
\h_k \= (1{-}\la^2)(x^1{-}a^1_k)+\im(1{-}\la^2)(x^2{-}a^2_k)-2\la(x^3{-}a^3_k)
\ =:\ \h-h_k
\end{equation}
restricted to a circle in $\C P^1$ 
(which does not pass through the zeros of any~$\h_k$)
and insert it into the matrix (\ref{fSU2}). 
Then solving the Riemann-Hilbert problem as in (\ref{fpm}) -- (\ref{+ps+})
we produce the field configuration (\ref{Aandp}) with
\begin{equation}\label{rn0} 
\r^{(n)}_0\ =\ \sum_{k=1}^n\, \frac{1}{2r_k} \= \sfrac12 \sum_{k=1}^n\, 
\bigl[(x^1{-}a^1_k)^2+(x^2{-}a^2_k)^2+(x^3{-}a^3_k)^2\bigr]^{-1/2}\ ,
\end{equation}
defined on $\R^3\backslash\{{\va}_1,\ldots,{\va}_n\}$. It describes a
Wu-Yang multi-monopole but not the most general one. This configuration
asymptotically approaches to the configuration (\ref{Ap}) describing 
$n$~Dirac monopoles.
One can see, for instance, that for $x^c\to a^c_k$ we have
\begin{equation}\label{Akphik}
A_a\quad\to\quad A_a^{(k)} \= \ve_{abc}\frac{\s_c}{2\im}\frac{x_k^b}{r^2_k}
\qquad \mbox{and}\qquad
\p\quad\to\quad\p^{(k)} \= \frac{\s_a}{2\im}\frac{x^a_k}{r^2_k}
\end{equation}
which can be transformed to the $k$th Dirac monopole with the help of 
matrices (\ref{gNgS}) where $x^a_k$ and $r_k$ are substituted for 
$x^a$ and $r$. Then, considering a small sphere $S^2_k$ surrounding
the point $x^c = a^c_k$, we easily find via~(\ref{mf}) that the flux 
through any such $S^2_k$ produces one unit of topological charge and their 
summation yields $n$ for the charge of the whole configuration.

\subsection{Noncommutative Wu-Yang monopoles}

\noindent
{\bf Generic form of the solution.}
Considering the noncommutative space $\R^3_\th$, we assume that
the $2\times 2$ matrix encoding noncommutative Wu-Yang U(2)
monopoles~\footnote{It is well known that in the noncommutative 
setup one should consider the group U(2) instead of SU(2) 
since the unit determinant condition is not preserved by 
noncommutative gauge transformations (see e.g.\cite{Matsubara}).} 
has the form (\ref{fSU2}) but with the holomorphic function $\r$
promoted to an operator $\hat\r$ acting in the one-oscillator Fock 
space $\Hcal$. We find that formulae (\ref{-ps-}) and (\ref{+ps+})
with $x^a\to\hat x^a$ solve the operator Riemann-Hilbert problem
\begin{equation}\label{fhSU2}
\fh_{+-}^{SU(2)}\  = \ 
\begin{pmatrix}
{\rh} &\la^{-1}\\-\la & 0
\end{pmatrix}
\  = \ \hat\psi^{-1}_+(\xh,\la)\,\hat\psi_-(\xh,\la)
\end{equation}
and eqs. (\ref{horho}) - (\ref{parr}) also hold their form for
$\rh =\rh_+ +\rh_0 +\rh_-$. Substituting (\ref{fhSU2}), (\ref{-ps-})
and (\ref{+ps+}) into formulae (\ref{Opflds}) and using the
recursion relations (\ref{parr}), we obtain
$$ 
\hat{A}_a\ =\ \ve_{abc}\frac{\s_c}{2\im} 
\Bigl(\rh_0^{\frac12}\pa_b\rh_0^{-\frac12}-
\rh_0^{-\frac12}\pa_b\rh_0^{\frac12}\Bigr)\ +\ \frac{{\mathbf 1}_2}{2} 
\Bigl(\rh_0^{-\frac12}\pa_b\rh_0^{\frac12}+
\rh_0^{\frac12}\pa_b\rh_0^{-\frac12}\Bigr) \ , 
$$ 
\begin{equation}\label{hatAp} 
\hat\p \ =\ \frac{\s_a}{2\im} 
\Bigl(\rh_0^{\frac12}\pa_a\rh_0^{-\frac12}-
\rh_0^{-\frac12}\pa_a\rh_0^{\frac12}\Bigr)\ . 
\end{equation} 
Here, the operator $\rh_0$ can be obtained from a given $\rh$ via the
Cauchy formula,
\begin{equation}\label{Cauchy}
\rh_0\ =\ \oint\limits_{|\la |=1}\!\frac{\diff\la}{2\pi\im\la}\,\rh\ .
\end{equation}
Recall that for the invertibility of the operator $\hat\h$ from
(\ref{hath}) we assume that $|\la |=1$ in (\ref{fhSU2})
and for $\rh$ (see section 4.2 for discussion of this).

\noindent
{\bf Noncommutative monopole solution.}
For the description of a noncommutative Wu-Yang monopole with moduli
$\vec{a}_k=(a^1_k, a^2_k, a^3_k)$ we take in (\ref{fhSU2})
\begin{equation}\label{rhk} 
\rh_{(k)}\ =\ \la\,\hat{\h}_k^{-1} \= 
\la\,(\yh_k - 2\la x^3_k - \la^2\hat{\yb}_k)^{-1}\ , 
\end{equation} 
where, as before,
\begin{equation}\label{yhk} 
\yh_k \= \yh - (a^1_k{+}\im a^2_k) \quad,\qquad
\hat{\yb}_k \= \hat{\yb} - (a^1_k{-}\im a^2_k)
\qquad\mbox{and}\qquad
x^3_k \= x^3 - a^3_k \ .
\end{equation} 
As usual (see e.g.~\cite{LP2} and references therein), we introduce a 
new basis
$\{|\ell\>_k,\ \ell=0,1,2,\ldots\}$ in $\Hcal$ via
\begin{equation}\label{yhklk} 
\yh_k|0\>_k \ =\ 0\quad,\qquad
|\ell\>_k\ =\ \frac{\hat{\yb}_k^\ell}{\sqrt{\ell !(2\th)^\ell}}|0\>_k
\quad,\qquad \ell =0,1,2,\ldots\quad .
\end{equation}
Then consider an operator (cf.~(\ref{solGN}))
\begin{equation}\label{solGNh} 
\xi_{(k)}\ :=\ (2{\th})^{1/4}\sum\limits_{\ell\ge 0} \xi_\ell (x^3_k) 
|\ell\>_k\< \ell|_k \qquad\mbox{with}\qquad  
\xi_\ell (x^3_k) \=\sqrt{ 
\frac{\Ups_{\ell-1}(\frac{x^3_k}{\sqrt{2\th}})} 
     {\Ups_{\ell}  (\frac{x^3_k}{\sqrt{2\th}})} 
} \ , 
\end{equation}
which solves the equation 
\begin{equation}\label{xik} 
\xi^2_{(k)} - \yh_k\,\xi^{-2}_{(k)}\,\hat{\yb}_k \= -2x^3_k \ . 
\end{equation} 
Using the operator (\ref{solGNh}) and formula (\ref{split3}), we obtain
\begin{align}\label{kCauchy}
\rh_{(k)0}&\= \oint\limits_{|\la|=1}\!\frac{\diff\la}{2\pi\im\la}\,\rh_{(k)} \=
\oint\limits_{|\la|=1}\!\frac{\diff\la}{2\pi\im}\,\hat{\h}_k^{-1} \=
\oint\limits_{|\la|=1}\!\frac{\diff\la}{2\pi\im}\,
(\xi_{(k)}^2 - \la\hat{\yb}_k)^{-1} 
(\la + \yh_k\xi^{-2}_{(k)})^{-1}
\nonumber\\ 
&\= \oint\limits_{|\la|=1}\!\frac{\diff\la}{2\pi\im}\,(2R_k)^{-1}\bigl[1 - 
(\la+\yh_k\xi^{-2}_{(k)})\,\hat{\yb}_k\,(2R_k)^{-1}\bigr]^{-1}
(\la+\yh_k\xi^{-2}_{(k)})^{-1}
\nonumber\\ 
&\=
(2R_k)^{-1}\oint\limits_{|\la|=1}\!\frac{\diff\la}{2\pi\im}\,
\left\{ (\la + \yh_k\xi^{-2}_{(k)})^{-1} + 
(\la+\yh_k\xi^{-2}_{(k)})\,\hat{\yb}_k\,(2R_k)^{-1}
(\la+\yh_k\xi^{-2}_{(k)})^{-1}+\ldots\right\}
\nonumber\\ 
&\=(2R_k)^{-1} \ ,
\end{align}
where
\begin{equation}\label{2Rk}
2R_k \ := \ \xi_{(k)}^2 + \yh_k\,\xi^{-2}_{(k)}\,\hat{\yb}_k 
\= 2\,(\xi_{(k)}^2 + x_k^3)
\end{equation} 
is an invertible operator as easily seen from (\ref{solGNh}). Note that in the 
commutative limit $R_k\to r_k=\sqrt{(x_k^1)^2+(x_k^2)^2+(x_k^3)^2}$ in
conformity with  (\ref{rl0}). Substituting (\ref{kCauchy}) into (\ref{hatAp}),
we obtain a noncommutative generalization of the Wu-Yang monopole.

\noindent
{\bf Noncommutative multi-monopoles.}
For obtaining noncommutative Wu-Yang multi-monopole solutions one may take
\begin{equation}\label{rh}
\rh^{(n)} \ =\ \sum\limits^n_{k=1}\rh_{(k)}
\end{equation} 
with $\rh_{(k)}$ from (\ref{rhk}) invertible for $|\la|{=}1$ and substitute it 
into (\ref{fhSU2}) to produce the multi-monopole matrix. 
This is nothing but the Atiyah-Ward~${\cal A}_1$ ansatz reduced
to three dimensions.
Then, (\ref{hatAp}) yields the multi-monopole configuration 
with the operator
\begin{equation}\label{rh0}
\rh_0^{(n)} \ =\ \sum^n_{k=1}\rh_{(k)0} \ =\ \sum^n_{k=1}(2R_k)^{-1}\ .
\end{equation} 
More general noncommutative Wu-Yang multi-monopoles may be inferred from
appendix~C.

\bigskip

\section{Nonabelian Bogomolny-Prasad-Sommerfield monopoles}

\subsection{BPS monopole on commutative $\R^3$}

\noindent
{\bf Matrix-valued holomorphic function.}
We come to the main topic of the paper, the construction of noncommutative
BPS monopoles via a Riemann-Hilbert problem. Let us first review the
commutative case. It is well known that
the spherically-symmetric SU(2) Bogomolny-Prasad-Sommerfield (BPS) 
monopole~\cite{Bogomolny, Prasad1, Prasad2} permits a twistor 
description~\cite{Ward1, Ward2, Corrigan2}.
Its central object is the holomorphic matrix-valued function
\begin{equation}\label{BPS}
f_{\rm BPS}\=
\begin{pmatrix}
(\e^w{-}\e^{-w})w^{-1} & \la^{-1} \e^{-w}\\[4pt] -\la\,\e^{-w} & w\,\e^{-w}
\end{pmatrix}
\end{equation} 
which defines the corresponding Riemann-Hilbert problem. 
Here we abbreviate by~$w$ the combination
\begin{equation}\label{w} 
w\ :=\ -\frac{\h}{\la} \= 2x^3+\la\yb-\la^{-1}y \=
(z+\la\yb)-(-\zb+\la^{-1}y) \ =:\ w_2-\tw_1 \ , 
\end{equation} 
where
\begin{equation}
w_2 \= z+\la\yb \qquad\textrm{and}\qquad \tw_1 \= -\zb+\la^{-1}y
\end{equation}
for fixed $y,z,\yb,\zb$ are holomorphic on $U_+$ and $U_-$, respectively.  
Note that $f_{\rm BPS}$ does not depend on $x^4$ and satisfies the reality  
condition  (\ref{Cond2}). 
 
It is convenient to factorize the matrix (\ref{BPS})
as follows (cf.~\cite{Ward1}):
\begin{align}\label{fBPS} 
f_{\rm BPS} &\= \begin{pmatrix}
(\e^{w_2}\e^{-\tw_1}{-}\e^{-w_2}\e^{\tw_1})w^{-1} & 
\la^{-1}\e^{-w_2}\e^{\tw_1} \\[4pt] 
-\la\,\e^{-w_2}\e^{\tw_1} & 
w\,\e^{-w_2}\e^{\tw_1}       
\end{pmatrix} \nonumber\\[8pt] 
&\= 
\begin{pmatrix} 0&\e^{-w_2}\\[4pt]-\e^{w_2}&\la w \e^{-w_2} \end{pmatrix} 
\begin{pmatrix} \la&0\\[4pt](\e^{2w_2}{-}\e^{2\tw_1})w^{-1}&\la^{-1}
\end{pmatrix}
\begin{pmatrix} \e^{-\tw_1}&0\\[4pt]0&\e^{\tw_1} \end{pmatrix} 
\nonumber\\[8pt]
&\= 
\begin{pmatrix} \e^{-w_2}&0\\[4pt] \la\,w\,\e^{-w_2}& \e^{w_2}\ \end{pmatrix}
\begin{pmatrix} w^{-1}(\e^{2w_2}{-}\e^{2\tw_1})&\la^{-1}\\[4pt]-\la&0
\end{pmatrix}
\begin{pmatrix} \e^{-\tw_1}&0\\[4pt]0&\e^{\tw_1} \end{pmatrix} \ . 
\end{align} 
The left and right factors are holomorphic on $U_+$ and $U_-$, respectively,
reducing our parametric Riemann-Hilbert problem to the middle factor, 
which takes the form
\begin{equation}\label{rho}
\begin{pmatrix} \r&\la^{-1}\\-\la&0 \end{pmatrix}
\qquad\mbox{with}\qquad
\r \= w^{-1}(\e^{2w_2} - \e^{2\tw_1})\ .
\end{equation}
However, this type of matrix has already been split in the previous section
(see also \cite{LP1}). Therefore, we can proceed like for the Wu-Yang SU(2)
monopole and expand $\r$ in a Laurent series in~$\la$,
\begin{equation}\label{sum13} 
\r \= \sum\limits^\infty_{m=-\infty}\r_m\,\la^m \= \r_-+\r_0 +\r_+\ , 
\end{equation} 
where 
\begin{equation} 
\r_+\ :=\ \sum\limits_{m >0}\r_m\,\la^m \qquad\mbox{and}\qquad  
\r_-\ :=\ \sum\limits_{m <0}\r_m\,\la^m \ . 
\end{equation} 
There are two important differences from the previous section, however.
First, our function~$\r$ given by~(\ref{rho}) does not satisfy the reality 
condition, i.e. $\r (x, -1/\lb ))^\+ \ne \r (x,\la)$. 
Second, it depends on~$x^4$, but in a simple way. We can make this explicit
by factorizing
\begin{equation}\label{rho1} 
\r \= \e^{-2\im x^4}\rt\ , 
\end{equation} 
where 
\begin{equation}\label{rho2} 
\rt\ :=\ \frac{1}{w}\,(\e^{2x^3}\e^{2\la\yb} - \e^{-2x^3}\e^{2\la^{-1}y})
\= \rt_+ + \rt_0 + \rt_-
\end{equation} 
is a function of~$\la$ and of $(x^1,x^2,x^3)$ only, 
with Laurent coefficients~$\rt_m$.

\noindent
{\bf Complex form of solution.}
Following the Wu-Yang case (cf.~(\ref{-ps-}) and (\ref{+ps+}))
but multiplying the additional holomorphic matrix factors from~(\ref{fBPS}),
we split
\begin{equation}\label{splitBPS}
f_{\rm BPS}\=\psi^{-1}_+\psi_-\ ,
\end{equation}
where now
\begin{equation}\label{ps-}
\psi_-=\r_0^{-\frac12}\!
\begin{pmatrix}\r_0{+}\rho_-&\la^{-1}\\[4pt] \la\rho_- & 1\end{pmatrix}\!\!
\begin{pmatrix}\e^{-\tw_1}&0\\[4pt] 0 & \e^{\tw_1}\end{pmatrix} ,\quad
\psi_-^{-1}=
\begin{pmatrix}\e^{\tw_1}&0\\[4pt] 0& \e^{-\tw_1}\end{pmatrix}\!\!
\begin{pmatrix} 1&-\la^{-1}\\[4pt] -\la\r_-& \r_0{+}\r_- \end{pmatrix}\!
\r_0^{-\frac12} ,
\end{equation}
\begin{equation}\label{ps+}
\psi_+=\r_0^{-\frac12}\!
\begin{pmatrix}1&\la^{-1}\rho_+\\[4pt] \la& \r_0{+}\rho_+\end{pmatrix}\!\!
\begin{pmatrix}\e^{w_2}&0\\[4pt] -\la \e^{-w_2}w&\e^{-w_2}\end{pmatrix} ,\quad
\psi_+^{-1}=
\begin{pmatrix}\e^{-w_2}&0\\[4pt] \la w \e^{-w_2}&\e^{w_2}\end{pmatrix}\!\!
\begin{pmatrix}\r_0{+}\rho_+&-\la^{-1}\rho_+\\[4pt] -\la& 1\end{pmatrix}\!
\r_0^{-\frac12} .
\end{equation}
{}Formulae (\ref{splitBPS}) -- (\ref{ps+}) solve a parametric Riemann-Hilbert 
problem on $\C P^1$ with external parameters $x\in\R^4$.

Having found $\psi_\pm (x,\la)$, one obtains a solution of the SDYM equations,
\begin{equation}\label{aaaa}
A_{\bar y}=\psi_+\pa_{\bar y}\psi_+^{-1}|_{\la =0}\ ,\quad
A_{\bar z}=\psi_+\pa_{\bar z}\psi_+^{-1}|_{\la =0}\ ,\quad
A_{y}=\psi_-\pa_{ y}\psi_-^{-1}|_{\la =\infty}\ ,\quad
A_{z}=\psi_-\pa_{z}\psi_-^{-1}|_{\la =\infty}\ .
\end{equation}
Substituting (\ref{ps-}) and (\ref{ps+}) into (\ref{aaaa}) and using the
recursion relations
\begin{equation}\label{rerel}
\pa_{\bar y}\,\r_{m+1}\ =\ \pa_z\,\r_m \qquad\mbox{and}\qquad
\pa_{\bar z}\,\r_{m+1}\ =\ -\pa_y\,\r_m
\end{equation}
implied by
\begin{equation}\label{holrho}
(\pa_{\bar y}-\la\pa_z)\r\ =\ (\pa_{\bar z}+\la\pa_y)\r\ =\ 0\ ,
\end{equation}
we get in real coordinates
\begin{equation}\label{aa}
A_a \= -\ve_{abc}\,\frac{\s_c}{2\im}\,\rt_0^{-1}\pa_b\rt_0 + \s_a
\qquad\mbox{and}\qquad
\p \= -\frac{\s_a}{2\im}\,\rt_0^{-1}\pa_a\rt_0 \ .
\end{equation}
One sees that the above configuration
$(A_a, \p )$ (derived for the first time by Manton~\cite{Manton}) does
not depend on $x^4$ (and therefore solves the Bogomolny equations) but
is not real since $A^\+_a\ne - A_a$. However, Manton has shown by direct
calculations that the solution (\ref{aa}) can be transformed by a complex
gauge transformation to a real form coinciding with the standard BPS
monopole~\cite{Manton} (see also~\cite{Adler, Rossi, Nahm2}).
The existence of such a gauge transformation follows from the fact that 
the matrix-valued function (\ref{BPS}) producing (\ref{aa}) satisfies 
the reality condition~(\ref{Cond2}).

\noindent
{\bf Nonunitary gauge transformation to a real form of solution.}
The matrix~$g$ producing the above-mentioned transformation
can be constructed as follows~(cf.~\cite{Crane}). 
{}From the reality condition (\ref{Cond2}) for the Birkhoff decomposition 
$f_{\rm BPS}=\psi^{-1}_+\psi_-$ we obtain
$$
\left(\psi_+^{-1}(x,-\bar\la^{-1})\ \psi_-(x,-\bar\la^{-1})\right)^\+ \=
\psi_+^{-1}(x, \la)\ \psi_-(x, \la) \qquad \Longrightarrow
$$
\begin{equation}\label{23}
\psi_+(x, \la)\ \psi_-^\+(x, -\bar\la^{-1})\=
\psi_-(x, \la)\ \psi_+^\+(x, -\bar\la^{-1})\ =:\ g^2\ .
\end{equation}
Note that, for any fixed $x\in\R^4$, the product 
$\psi_+(x,\la)\psi_-^\+(x,-\bar\la^{-1})$ is holomorphic (in $\la$) 
on $U_+$ and $\psi_-(x,\la)\psi_+^\+(x,-\bar\la^{-1})$ is holomorphic on $U_-$.
Therefore, $g^2$ does not depend on $\la$ due to Liouville's theorem
(globality on $\C P^1=U_+\cup U_-$). Moreover, we have
\begin{equation}\label{24}
\psi_+^\+(x,-\bar\la^{-1}) \= \psi_-^{-1}(x,\la)\,g^{2} \=
\psi_+^\+(x,-\bar\la^{-1})\,(g^\+)^{-2} g^2
\qquad\Longrightarrow\qquad g^2\=(g^\+)^2\ .
\end{equation}
If $g^2$ has no negative eigenvalues, it follows that $g=g^\+$,
i.e. $g=g(x)$ is a hermitean matrix. 
One can calculate it for any~$\la$ in~(\ref{23}), 
e.g. by taking the square root of
\begin{equation}\label{g^2}
g^2\=\psi_+(x,\la)\,\psi_-^\+(x,-\bar\la^{-1})|_{\la =0} \ .
\end{equation}

Substituting $\psi_\pm$ from (\ref{ps-}) and (\ref{ps+}) into (\ref{g^2}), 
we obtain
\begin{align}\label{g-2}
g^{-2}&=\ (\psi_-^\+(x,-\bar\la^{-1} ))^{-1}\psi_+^{-1}(x,\la)|_{\la=0}\ =\
(\psi_+^\+(x,-\bar\la^{-1} ))^{-1}\psi_-^{-1}(x,\la)|_{\la=\infty}
\nonumber\\[8pt]
&=\ (\rt_0^\+)^{-\frac{1}{2}}
\begin{pmatrix}
\rt_0\e^{-2x^3}+\rt_{-1}^\+ y\rt_{0}\quad & \quad
-\rt_1\e^{-2x^3}-\rt_{-1}^\+ \e^{2x^3}-\rt_{-1}^\+ y\rt_{1} \\[4pt]
-\rt_0^\+y\rt_0 & \rt_0 \e^{2x^3}+\rt_0^\+y\rt_{1}
\end{pmatrix}
\,\rt_0^{-\frac{1}{2}} \nonumber\\[8pt]
&=\ (\rt_0^\+)^{-\frac{1}{2}}
\begin{pmatrix}
\rt_0^\+ \e^{-2x^3}+\rt_{0}^\+\bar y\rt_{-1} &
-\tilde\r_0\bar y\tilde\r_0^\+ \\[4pt]
-\rt_1^\+ \e^{-2x^3}- \rt_{-1} \e^{2x^3}-\rt_{1}^\+\bar y\rt_{-1}\quad & \quad
\rt_0^\+ \e^{2x^3}+\rt_1^\+\bar y\rt_0
\end{pmatrix}
\,\rt_0^{-\frac{1}{2}}\ ,
\end{align}
which is $x^4$-independent as it should be.

After having taken the square root of (\ref{g-2}) one can introduce
\begin{equation}\label{31}
\psi_-^g\ :=\ g^{-1}\psi_- \qquad\mbox{and}\qquad \psi_+^g\ :=\ g^{-1}\psi_+\ ,
\end{equation}
which satisfy the reality condition (\ref{Cond1}),
\begin{equation}\label{rea}
\left(\psi_+^g(x,-{\bar\la}^{-1})\right)^\+ \=
\psi_+^\+(x,-{\bar\la}^{-1})\,g^{-1} \= \psi_-^{-1}(x,\la)\,g^2g^{-1} \=
\left(\psi_-^g(x,\la)\right)^{-1}
\end{equation}
due to eq.~(\ref{24}). By direct calculation one can show that the
gauge-transformed fields
\begin{equation}\label{Aga}
A^g_a\=g^{-1} A_a g+g^{-1}\pa_a g \qquad\mbox{and}\qquad \p^g\=g^{-1}\p\,g
\end{equation}
are indeed antihermitean and thus represent a real form of our solution
to the Bogomolny equations.

\noindent
{\bf Explicit solution.}
It remains to compute the Laurent coefficients $\rt_0$ and $\rt_{\pm1}$ of the 
function $\rt=w^{-1}(\e^{2x^3}\e^{2\la\yb}{-}\e^{-2x^3}\e^{2\la^{-1}y})$
which enter the expressions (\ref{aa}) and~(\ref{g-2}). 
A point of concern may be that the denominator 
\begin{equation}\label{dnmntr} 
w\=w_2-\tw_1\=\la\yb+2 x^3 -\la^{-1}y 
\= \frac{1}{\la}\,(\la-\la_1)\,(\la-\la_2)\,\yb
\end{equation} 
of $\rt$ vanishes at points $\la_1, \la_2\in \C P^1$ given by formulae 
(\ref{l1l2}). Nevertheless, the function $\rt$ is well defined at these  
points (and has poles only at $\la =0$ and $\la =\infty$) as can be seen  
from its Laurent-series expansion. 

{}From the experience with the Dirac and Wu-Yang monopoles we expect that
the Riemann-Hilbert problem for the noncommutative BPS monopole, to be treated 
in the next subsection, takes the same form as eq.~(\ref{fBPS})
modulo ordering ambiguities. Anticipating that the noncommutative cousin 
of~$\r$ will appear in Weyl-ordered form, we shall organize the following 
(commutative) calculation in a symmetric manner which will remain valid in the
noncommutative case (with coordinates promoted to operators).
We begin by rewriting
\begin{align}\label{rsym} \nonumber
\r &\= \e^{w_2}\,w^{-1}\,\e^{w_2}\ -\ \e^{\tw_1}\,w^{-1}\,\e^{\tw_1} \\[4pt]
   &\= \e^{z+\la\yb}\la\,\yb^{-1}(\la{-}\la_2)^{-1}(\la{-}\la_1)^{-1}
       \e^{z+\la\yb}\ -\ \e^{-z+\la^{-1}y}\la\,\yb^{-1}(\la{-}\la_2)^{-1}
       (\la{-}\la_1)^{-1}\e^{-z+\la^{-1}y}\ .
\end{align}
Making use of the relations
\begin{equation}
w_2(\la_1)\=\tw_1(\la_1) \qquad\textrm{and}\qquad w_2(\la_2)\=\tw_1(\la_2)
\end{equation}
we decompose $\r$ into
\begin{align}\label{r+} \nonumber
\r_+ &\= 
 \e^{w_2(\la)}\la\,\yb^{-1}(\la {-}\la_2)^{-1}(\la {-}\la_1)^{-1}\e^{w_2(\la)}
\\[4pt] &\
-\e^{w_2(\la_1)}\la\,\yb^{-1}(\la_1{-}\la_2)^{-1}
(\la {-}\la_1)^{-1}\e^{w_2(\la_1)} -\e^{w_2(\la_2)}\la\,\yb^{-1}
(\la {-}\la_2)^{-1}(\la_2{-}\la_1)^{-1}\e^{w_2(\la_2)} \ ,\phantom{x}
\end{align}
\begin{align}\label{r-} \nonumber
\r_- &\=
-\e^{\tw_1(\la)}\la\,\yb^{-1}(\la{-}\la_2)^{-1}
(\la{-}\la_1)^{-1}\e^{\tw_1(\la)}
\\[4pt] &\
+\e^{\tw_1(\la_1)}\yb^{-1}(\la_1{-}\la_2)^{-1}
(\la{-}\la_1)^{-1}\la_1\e^{\tw_1(\la_1)}
+\e^{\tw_1(\la_2)}\yb^{-1}\la_2(\la{-}\la_2)^{-1}
(\la_2{-}\la_1)^{-1}\e^{\tw_1(\la_2)} \ ,
\end{align}
\begin{align}\label{rnull} 
\r_0 \= 
 \e^{\tw_1(\la_1)}\yb^{-1}(\la_1{-}\la_2)^{-1}\e^{\tw_1(\la_1)}
+\e^{w_2(\la_2)}  \yb^{-1}(\la_2{-}\la_1)^{-1}\e^{w_2(\la_2)} \ .
\phantom{XXXXXXXXXXf}
\end{align}
Solving (\ref{dnmntr}) for $\la_1$ and $\la_2$ as in~(\ref{l1l2}) we obtain 
(for commuting quantities)
\begin{equation}\label{zeros}
\la_1 \=-\frac{r+x^3}{\yb}\=-\frac{y}{r-x^3}
\qquad\textrm{and}\qquad
\la_2 \=\frac{y}{r+x^3}=\frac{r-x^3}{\yb}
\end{equation}
with $\ r^2=y\yb+x^3 x^3\ $ and therefore
\begin{equation}
(\la_1{-}\la_2)\yb\=-2r\ ,\qquad
\tw_1(\la_1)\=-\im x^4-r\ ,\qquad
w_2(\la_2)  \=-\im x^4+r
\end{equation}
which simplifies (\ref{rnull}) to
\begin{equation}\label{rnull2}
\r_0 \=
\e^{-\im x^4-r}(-2r)^{-1}\e^{-\im x^4-r} + 
\e^{-\im x^4+r}( 2r)^{-1}\e^{-\im x^4+r}
\end{equation}
yielding
\begin{equation}\label{rt0}
\rt_0 \= \e^{2\im x^4}\,\r_0 \= \frac{\sinh (2r)}{r}\ .
\end{equation}
Likewise, from (\ref{r+}) and (\ref{r-}) one can easily obtain 
\begin{align}\label{rt+} 
\rt_1 \= \e^{2\im x^4} \lim\limits_{\la\to 0}(\la^{-1}\r_+)
&\= -\frac{\e^{2x^3}}{y} 
+\frac{(r{-}x^3)}{y}\frac{\e^{-2r}}{2r} 
+\frac{(r{+}x^3)}{y}\frac{\e^{2r}}{2r} \ , \\[8pt]
\label{rt-}
\rt_{-1} \= \e^{2\im x^4} \lim\limits_{\la\to\infty}(\la\,\r_-)
&\= -\frac{\e^{-2x^3}}{\yb} 
+\frac{(r{+}x^3)}{\yb}\frac{\e^{-2r}}{2r} 
+\frac{(r{-}x^3)}{\yb}\frac{\e^{2r}}{2r} \ . 
\end{align} 

The simple form of~$\rt_0$ immediately yields the complex form of the solution,
\begin{equation}
A_a \= \ve_{abc}\,\frac{\s_c}{2\im}\,\frac{x_b}{r}
\Bigl(\frac{1}{r}-2\,{\rm coth}(2r)\Bigr) + \s_a
\qquad\mbox{and}\qquad
\p \= \frac{\s_a}{2\im}\,\frac{x_a}{r}
\Bigl(\frac{1}{r}-2\,{\rm coth}(2r)\Bigr)\ .
\end{equation}
Next, we evaluate the nonunitary gauge matrix~$g$.
Inserting the functions $\rt_0$ 
and $\rt_{\pm 1}$ into (\ref{g-2}) simplifies it to 
\begin{equation}\label{g2e} 
g^2\=\e^{2x^a\s_a} \qquad\mbox{which allows for}\qquad g=\e^{x^a\s_a}\  . 
\end{equation} 
Since this $g$ commutes with~$\p$ we have $\p^g=\p$, 
but the gauge potential~$A_a$ transforms to an antihermitean~$A_a^g$. 
The real form of the solution, 
\begin{equation}\label{Aa}
A^g_a\= \ve_{abc}\,\frac{\s_c}{2\im}\,\frac{x_b}{r}
\Bigl(\frac{1}{r}-\frac{2}{\sinh(2r)}\Bigr)
\qquad\mbox{and}\qquad
\p^g \= \frac{\s_a}{2\im}\,\frac{x_a}{r}
\Bigl(\frac{1}{r}-2\,{\rm coth}(2r)\Bigr)
\end{equation}
indeed coincides with the BPS monopole in its standard form.

\subsection{Noncommutative BPS monopole} 
 
\noindent
{\bf Some formulae.}
We are now ready to generalize the framework of the previous subsection
to the noncommutative case.
For the noncommutative space $\R^3_\th$, we assume that a matrix-valued
function encoding a charge one noncommutative U(2) BPS monopole still 
has the form (\ref{BPS}) but with $w$ being considered as an  
operator~$\wh$ acting in the one-oscillator Fock space $\Hcal$. So, we define
\begin{equation}
\wh_2\= z + \la\hat{\yb} \qquad\textrm{and}\qquad 
\hat{\tw}_1\= \la^{-1}\yh - \zb
\end{equation}
and
\begin{equation}\label{wh}  
\wh\ :=\ \wh_2-\hat{\tw}_1 \= z + \zb + \la\hat{\yb} - \la^{-1}\yh \=  
2x^3 + \la\hat{\yb} - \la^{-1}\yh\ . 
\end{equation} 
The commutation relations 
\begin{equation} \label{ncco1} 
[\,{\yh}\,,\,{\hat{\yb}}\,]\ =\ 2\th \qquad\textrm{and}\qquad 
\textrm{other commutators}\ =\ 0 \qquad\Longrightarrow\qquad
[\hat{\tw}_1, \wh_2] \= 2\th
\end{equation} 
imply that
\begin{equation}  
\wh_2\,\e^{\a\hat{\tw}_1} \= \e^{\a\hat{\tw}_1}(\wh_2-2\a\th )
\qquad\textrm{and}\qquad
\hat{\tw}_1\,\e^{\a \wh_2} \= \e^{\a \wh_2}(\hat{\tw}_1+2\a\th )\ ,
\end{equation} 
\begin{equation}  
\wh\,\e^{\a\hat{\tw}_1} \= \e^{\a\hat{\tw}_1}(\wh -2\a\th )
\qquad\textrm{and}\qquad
\wh\,\e^{\a \wh_2} \= \e^{\a \wh_2}(\wh -2\a\th )\ ,
\end{equation} 
\begin{equation}\label{ew}  
\e^{\pm\wh} \= \e^{-{\th}}\,\e^{\pm{\wh}_2}\,\e^{\mp{\hat{\tw}}_1}
\= \e^{\th}\,\e^{\mp\hat{\tw}_1}\,\e^{\pm{\wh}_2} \ ,
\end{equation} 
where $\a$ is an arbitrary constant. Also we have 
\begin{equation} 
(\hat{\tw}_1(-\bar\la^{-1}))^\+ = -\wh_2(\la)\ , \qquad
(\wh_2(-\bar\la^{-1}))^\+ = -\hat{\tw}_1(\la) \qquad\Longrightarrow\qquad
(\wh(-\bar\la^{-1}))^\+ = \wh(\la)\ .
\end{equation} 
 
\noindent
{\bf Matrix-valued operator.} 
Repeating arguments from section 4.2, we admit in (\ref{wh}) only 
such $\la$ that the operator $\wh$ will not have zero modes on the 
Hilbert spaces $\Hcal$ or $\Hcal^*$. As was shown in section 4.2, the
operator $\wh$ is invertible iff $|\la |=1$. In the following we consider 
the operator-valued matrix 
\begin{equation}\label{hBPS} 
{\fh}_{\rm BPS}\= 
\begin{pmatrix} 
(\e^{\wh}{-}\e^{-\wh}){\wh}^{-1}&\la^{-1} \e^{-\wh}\\ 
-\la\,\e^{-\wh} &{\wh}\,\e^{-\wh} 
\end{pmatrix} 
\end{equation} 
restricted to $S^1=\{\la\in\C P^1: |\la |=1\}$,
where $\yh$, $\hat\yb$ and $x^3$ are considered as external ``parameters''.
The calculations in the noncommutative case can almost literally be copied from
the commutative situation treated in the previous subsection. 
 
\noindent
{\bf Splitting.}
The operator-valued matrix (\ref{hBPS}) is split as follows, 
\begin{align}\label{eBPS} 
\e^{-\th}\fh_{\rm BPS} & 
\=\begin{pmatrix}
(\e^{\wh_2}\e^{-\hat{\tw}_1}{-}\e^{-\wh_2}\e^{\hat{\tw}_1})\wh^{-1} & 
\la^{-1}\e^{-\wh_2} \e^{\hat{\tw}_1} \\[4pt] 
-\la\,\e^{-\wh_2} \e^{\hat{\tw}_1} &
\wh\,\e^{-\wh_2} \e^{\hat{\tw}_1}  
\end{pmatrix} \nonumber\\[8pt] 
&\=\begin{pmatrix}
0 & \e^{-\wh_2} \\[4pt] -\e^{\wh_2} & \la \wh \e^{-\wh_2} 
\end{pmatrix}  
\begin{pmatrix}
\la & 0 \\[4pt] (\e^{2\wh_2}{-}\e^{2\hat{\tw}_1})(\wh{-}2\th)^{-1} & \la^{-1} 
\end{pmatrix}  
\begin{pmatrix}
\e^{-\hat{\tw}_1} & 0 \\[4pt] 0 & \e^{\hat{\tw}_1}
\end{pmatrix} \nonumber\\[8pt] 
&\=\begin{pmatrix}
\e^{-\wh_2} & 0 \\[4pt] \la\,\wh\,\e^{-\wh_2} & \e^{\wh_2}
\end{pmatrix}  
\begin{pmatrix}
(\wh{+}2\th)^{-1}(\e^{2\wh_2}{-}\e^{2\hat{\tw}_1}) & \la^{-1} \\[4pt] -\la & 0
\end{pmatrix}  
\begin{pmatrix}
\e^{-\hat{\tw}_1} & 0 \\[4pt] 0 & \e^{\hat{\tw}_1}
\end{pmatrix} \nonumber\\[8pt] 
&\= \ph_+^{-1}\,\ph_-\ , 
\end{align} 
where 
\begin{equation}
\ph_-=\rh_0^{-\frac12}\!
\begin{pmatrix}\rh_0{+}\rh_-&\la^{-1}\\[4pt] \la\rh_- & 1\end{pmatrix}\!\!
\begin{pmatrix}\e^{-\hat\tw_1}&0\\[4pt] 0 & \e^{\hat\tw_1}\end{pmatrix} ,
\quad
\ph_-^{-1}=
\begin{pmatrix}\e^{\hat\tw_1}&0\\[4pt] 0& \e^{-\hat\tw_1}\end{pmatrix}\!\!
\begin{pmatrix} 1&-\la^{-1}\\[4pt] -\la\rh_-& \rh_0{+}\rh_- \end{pmatrix}\!
\rh_0^{-\frac12} ,
\end{equation}
\begin{equation}
\ph_+=\rh_0^{-\frac12}\!
\begin{pmatrix}1&\la^{-1}\rh_+\\[4pt] \la& \rh_0{+}\rh_+\end{pmatrix}\!\!
\begin{pmatrix}\e^{\wh_2}&0\\[4pt]-\la\e^{-\wh_2}\wh&\e^{-\wh_2}\end{pmatrix},
\quad
\ph_+^{-1}=
\begin{pmatrix}\e^{-\wh_2}&0\\[4pt]\la\wh\e^{-\wh_2}&\e^{\wh_2}\end{pmatrix}\!\!
\begin{pmatrix}\rh_0{+}\rh_+&-\la^{-1}\rh_+\\[4pt] -\la& 1\end{pmatrix}\!
\rh_0^{-\frac12} 
\end{equation}
take the same form as in the commutative case (cf.~(\ref{ps-}) and (\ref{ps+})),
and 
\begin{align} \nonumber \label{r+0-} 
\rh\ :=\ &(\wh{+}2\th)^{-1} (\e^{2\wh_2}- \e^{2\hat{\tw}_1}) 
\= \e^{\wh_2}{\wh^{-1}}\e^{\wh_2} 
 - \e^{\hat{\tw}_1}{\wh^{-1}}\e^{\hat{\tw}_1}\\[4pt] \nonumber
\= &\int_{-1}^1 \!\diff t \ \e^{(1+t)\wh_2 + (1-t)\hat\tw_1}
\= \e^{2\im x^4} \int_{-1}^1 \!\diff t \ \e^{2tx^3}\,
   \e^{(1+t)\la\hat\yb + (1-t)\la^{-1}\yh} \\[4pt]
\= &\rh_+ +\rh_0 +\rh_- 
\= \e^{-2\im x^4}\,(\hat\rt_+ +\hat\rt_0 +\hat\rt_-)
\= \e^{-2\im x^4}\,\hat\rt
\end{align} 
is indeed Weyl ordered.
Notice that also the operators $\wh\pm 2\th$ have no zero modes 
on $\Hcal$ or $\Hcal^*$ if $|\la |=1$. 

\noindent
{\bf Explicit form of $\hat\rt_0$ and $\hat\rt_{\pm1}$.}
For the computation of $\hat\rt$ we may go back to the commutative case
and simply put hats in eqs.~(\ref{rsym})--(\ref{rnull}), including over
$\la_1$ and $\la_2$ which implicitly depend on $\hat{y}$ and $\hat\yb$.
However, to proceed from the noncommutative extension of~(\ref{rnull})
one cannot employ the expression~(\ref{zeros}) for the zeros 
$\hat\la_1$ and $\hat\la_2$ of $\wh$.
Instead, we use the factorization~(\ref{split3}) introduced for the
noncommutative Dirac monopole,
\begin{equation}
\wh \= -(\xi +\la^{-1}\hat{y} \xi^{-1})\,(\xi -\la \xi^{-1}\hat{\yb})
\qquad\textrm{with}\qquad
\xi^2 - \yh\xi^{-2}\hat{\yb} \= -2x^3 \ ,
\end{equation}
which involves the hermitean operator~$\xi$.
Comparison with~(\ref{dnmntr}) implies~\footnote{
We choose $\hat\la_2|0\>=0$.}
\begin{equation}
\hat\la_1 \= -\yh\,\xi^{-2} \qquad\textrm{and}\qquad 
\hat\la_2\,\hat\yb \= \xi^2 \qquad\textrm{hence}\qquad
\hat\la_2 \= \xi^2\,(\yh\hat\yb)^{-1} \yh \ .
\end{equation}
Defining again $R:=x^3+\xi^2$, we may write
\begin{equation}
(\hat\la_1{-}\hat\la_2) \= -2 R\,(\yh\hat\yb)^{-1} \yh \qquad\textrm{and}\qquad
\hat\la_1\,\hat\la_2    \= - \yh\,(\yh\hat\yb)^{-1} \yh
\end{equation}
and also
\begin{equation}
\hat\tw_1(\hat\la_1)\= \wh_2(\hat\la_1) \= -\im x^4 -R \qquad\textrm{and}\qquad
\hat\tw_1(\hat\la_2)\= \wh_2(\hat\la_2) \= -\im x^4 +R \ .
\end{equation}
Employing these relations in (the noncommutative version of) (\ref{rnull}) 
we get
\begin{equation}
\rh_0 \= \e^{-\im x^4-R}(-2R)^{-1}\e^{-\im x^4-R}
       + \e^{-\im x^4+R}( 2R)^{-1}\e^{-\im x^4+R}
\end{equation}
which means
\begin{equation}\label{rt0h}
\hat\rt_0 \= \frac{\sinh (2R)}{R} \= \hat\rt^\+_0\ .
\end{equation}

Analogously, with the help of a noncommutative variant of~(\ref{zeros}),
\begin{equation}
\yh\,(R{-}x^3)^{-1}    \= (R{+}x^3)\,(\yh\hat\yb)^{-1} \yh
\qquad\textrm{and}\qquad
(R{-}x^3)^{-1} \hat\yb \= \hat\yb\,(\yh\hat\yb)^{-1} (R{+}x^3) \ ,
\end{equation}
and the relations
\begin{equation}
1 \= \yh\,\hat\yb\,(\yh\hat\yb)^{-1} \=
\yh\,(\hat\yb\yh)^{-1}\,\hat\yb \= (\yh\hat\yb)^{-1}\,\yh\,\hat\yb
\end{equation}
we extract from (\ref{r+}) and (\ref{r-}) the Laurent coefficients
\begin{align} \label{rt+h}
\hat{\rt}_1 &\= -\hat\yb\,(\yh\hat\yb)^{-1} \e^{2x^3}
+(2R)^{-1} (R{-}x^3)\,\e^{-R} \hat\yb\,(\yh\hat\yb)^{-1} \e^{-R}
+\e^R\,\hat\yb\,(\yh\hat\yb)^{-1} \e^R (R{+}x^3) (2R)^{-1}\ , \\[8pt]
\label{rt-h}
\hat{\rt}_{-1} &\= -(\yh\hat\yb)^{-1} \yh\,\e^{-2x^3}
+(2R)^{-1} (R{+}x^3)\,\e^{-R} (\yh\hat\yb)^{-1} \yh\,\e^{-R}
+\e^R (\yh\hat\yb)^{-1} \yh\,\e^R (R{-}x^3) (2R)^{-1} \ .
\end{align}
We notice a formal similarity with the commutative result 
(\ref{rt0})--(\ref{rt-}),
which becomes concrete in the commutative limit (cf.~(\ref{xi+})),
\begin{equation}
\xi^2 \ \to\ \xi_+^2 \= r-x^3 \qquad\textrm{hence}\qquad R\ \to\ r\ .
\end{equation}

\noindent
{\bf Noncommutative BPS monopole in a complex gauge.}
Substituting (\ref{rt0h}) into (\ref{Opflds}) and using the 
recursion relations~\footnote{These recursion relations keep their form 
in the noncommutative case.} 
(\ref{rerel}) for the Laurent coefficient $\hat\r_m$ of $\hat\r$, 
we get (cf.~eq.~(3.20) in~\cite{LP1})  
\begin{align} 
\hat A_\mu\ &=\ \bar\eta^a_{\mu\nu}\,\frac{\s_a}{2\im} 
\Bigl(\rh_0^{\frac12}\pa_\nu\rh_0^{-\frac12}-
\rh_0^{-\frac12}\pa_\nu\rh_0^{\frac12}\Bigr) 
+ \frac{{\mathbf 1}_2}{2} 
\Bigl(\rh_0^{-\frac12}\pa_\mu\rh_0^{\frac12}+
\rh_0^{\frac12}\pa_\mu\rh_0^{-\frac12}\Bigr) \\[8pt] 
\nonumber \Longrightarrow\quad
\hat A_a\ &=\ \ve_{abc}\,\frac{\s_c}{2\im} 
\Bigl(\hat\rt_0^{\frac12}\pa_b\hat\rt_0^{-\frac12}-
\hat\rt_0^{-\frac12}\pa_b\hat\rt_0^{\frac12}\Bigr) 
+ \frac{{\mathbf 1}_2}{2} 
\Bigl(\hat\rt_0^{-\frac12}\pa_a\hat\rt_0^{\frac12}+
\hat\rt_0^{\frac12}\pa_a\hat\rt_0^{-\frac12}\Bigr) 
+\s_a\ , \\[4pt] \label{Amu4}
\hat\p\ &\equiv\ \hat A_4\ =\ \frac{\s_a}{2\im} 
\Bigl(\hat\rt_0^{\frac12}\pa_a\hat\rt_0^{-\frac12}-
\hat\rt_0^{-\frac12}\pa_a\hat\rt_0^{\frac12}\Bigr)\ , 
\end{align} 
where the operator $\hat\rt_0$ is given by (\ref{rt0h}). It is interesting that
$\hat\rt_0$ is expressed via the operator $\xi^2$ which defines the 
noncommutative Dirac monopole (see section 4.2). One sees that 
the configuration $(\hat A_a,\hat\p)$ does not depend on $x^4$.  We have 
obtained a solution of the noncommutative Bogomolny equations which is not 
real since $\hat A^\+_a\ne -\hat A_a$ analogously to the commutative case. 
Thus, (\ref{Amu4}) is the noncommutative BPS monopole written 
in a complex gauge.\footnote{
Note, however, that $\hat\p$ is already antihermitean.} 
 
\noindent
{\bf Gauge transformation to a real configuration.}
Similar to the commutative case one can derive an operator-valued matrix 
$\hat g(\xh)$ which generates a gauge transformation to a real form of the 
noncommutative BPS monopole. In fact, the formulae (\ref{23}) -- (\ref{g^2}) 
for the above-mentioned matrix and the gauge transformations (\ref{31}) and 
(\ref{Aga}) apply literally to the noncommutative setup.
The explicit form of $\hat g^{-2}$ in terms of $\hat\rt_0$ and $\hat\rt_{\pm1}$ 
is, however, more involved than in the commutative situation:
\begin{align}\label{g-2h} 
\hat g^{-2}\ &=\ 
(\ph_-^\+(\xh ,-\bar\la^{-1} ))^{-1}\,\ph_+^{-1}(\xh ,\la)|_{\la =0}\ =\  
(\ph_+^\+(\xh ,-\bar\la^{-1} ))^{-1}\,\ph_-^{-1}(\xh ,\la)|_{\la =\infty}
\nonumber\\[4pt] 
&=\ (\hat\rt_0^\+)^{-\frac{1}{2}} 
\begin{pmatrix} 
\hat\rt_0\e^{-2x^3}+\hat\rt_{-1}^\+ \yh\hat\rt_{0}\quad &\quad 
-\hat\rt_1\e^{-2x^3}-\hat\rt_{-1}^\+ \e^{2x^3}-\hat\rt_{-1}^\+ \yh\hat\rt_{1}
\\[4pt] 
-\hat\rt_0^\+\yh\hat\rt_0& \hat\rt_0 \e^{2x^3}+\hat\rt_0^\+\yh\hat\rt_{1} 
\end{pmatrix} 
\hat\rt_0^{-\frac{1}{2}} \nonumber\\[8pt] 
&=\ (\hat\rt_0^\+)^{-\frac{1}{2}} 
\begin{pmatrix} 
\hat\rt_0^\+ \e^{-2x^3}+\hat\rt_{0}^\+\hat{\yb}\hat\rt_{-1} &  
-\hat\rt_0\hat{\yb}\hat\rt_0^\+ 
\\[4pt] 
-\hat\rt_1^\+ \e^{-2x^3}-\hat\rt_{-1}\e^{2x^3}
  -\hat\rt_{1}^\+\hat{\yb}\hat\rt_{-1}\quad &\quad 
\hat\rt_0^\+ \e^{2x^3}+\hat\rt_1^\+\hat{\yb}\hat\rt_0 
\end{pmatrix} 
\hat\rt_0^{-\frac{1}{2}} \nonumber\\[8pt] 
&=\ (\hat\rt_0)^{-\frac{1}{2}} 
\begin{pmatrix} 
\hat\rt_0(\e^{-2x^3}+\hat{\yb}\hat\rt_{-1}) &  
-\hat\rt_0\hat{\yb}\hat\rt_0 \\[4pt] 
-\hat\rt_0{\yh}\hat\rt_0 &  
\hat\rt_0 (\e^{2x^3}+\hat{y}\hat\rt_{-1}) 
\end{pmatrix} 
\hat\rt_0^{-\frac{1}{2}} \ , 
\end{align} 
where we used the hermiticity of $\hat\rt_0$ as given by (\ref{rt0h}). 
After inserting the expressions (\ref{rt0h}), (\ref{rt+h}) and (\ref{rt-h})
for $\hat\rt_0$, $\hat\rt_1$ and $\hat\rt_{-1}$, respectively, it is unlikely
that the matrix (\ref{g-2h}) can be converted
into a nice exponential form (cf.~(\ref{g2e}) in the commutative case) 
which would permit us to explicitly take the square root $\hat g$ of $\hat g^2$
and insert into the formulae (\ref{31}) and (\ref{Aga}). Therefore, we must
be content with an implicit real form of the noncommutative BPS monopole,
given by (\ref{Amu4}) and~(\ref{g-2h}). 

Noncommutative BPS multi-monopoles may be treated in the same spirit,
by employing matrices more general than~(\ref{hBPS}) which are known
for the commutative case~\cite{Ward2,Corrigan2}.

\bigskip

\section{Conclusions} 

\noindent
In this paper we have developed a unified approach to noncommutative monopoles
of Dirac, Wu-Yang and BPS type, by reformulating their construction as
a parametric Riemann-Hilbert problem. Although this method is well known
in the commutative situation, its extension to the noncommutative case
seems to be new, and it is equally fruitful.~\footnote{
It works even in the framework of string field theory~\cite{LPLPU}.}
In fact, most steps -- writing the Bogomolny equation, formulating the
auxiliary linear system, posing the Riemann-Hilbert problem for a suitably
chosen matrix-valued function and even splitting the latter -- 
can be generalized without much difficulty.
This was demonstrated here by treating the commutative and noncommutative
construction in succession for each type of monopole.
In particular, the noncommutative Dirac and U(2) BPS monopoles were rederived,
and explicit noncommutative U(2) Wu-Yang multi-monopoles were constructed
for the first time.

Compared to previous derivations of noncommutative monopole configurations,
our presentation is just as practical in calculations as it is systematic.  
For instance, the construction of the 
noncommutative Dirac monopole~\cite{Gross:2000wc} reduces to a few lines
of algebra, and the key relation (see~(\ref{key}) below) emerges naturally. 
Of course, some of the computational details are more
intricate than in the commutative case. Let us mention two points here
for the example of the noncommutative U(2) BPS monopole.
First, the Riemann-Hilbert problem generically yields the monopole potential
in a complex gauge 
(which we constructed explicitly even in the noncommutative case), 
but it also provides one with a recipe for gauge 
transforming to a real configuration. Yet, the requisite nonunitary gauge
transformation seems to simplify only in the commutative limit. Second, 
it is not the factorization of the $2{\times}2$ matrix~$\fh_{\rm{BPS}}$ 
but the decomposition of its nontrivial entry~$\rh$ which requires more skill 
in the noncommutative setup. 

As a technical remark, in all cases considered in this paper we are led to
consider for fixed~$x$ the zeros $\la_1$ and $\la_2$ of the function
\begin{equation}
\h(x,\la) \=y - 2\la x^3 - \la^2\yb \=
(\la\xi_+ + y\xi_+^{-1})\,(\xi_+ -\la \xi_+^{-1}\yb)
\end{equation}
which gets deformed noncommutatively to
\begin{equation}
\hat\h(\xh,\la) \= \yh - 2\la x^3 - \la^2\hat{\yb}\=
(\la\xi + \hat{y} \xi^{-1})\,(\xi -\la \xi^{-1}\hat{\yb}) \ .
\end{equation}
Here, the operator $\xi$ is defined by 
\begin{equation} \label{key}
\xi^2-\yh\xi^{-2}\hat\yb+2x^3 \=0
\end{equation}
and in the commutative limit tends to the function
\begin{equation}
\xi_+ \= (r-x^3)^{1/2} \qquad\textrm{with}\quad
r^2\=y\yb+x^3 x^3 \ .
\end{equation}
Since the deformed variable $\hat\h$ still commutes with~$\la$, 
the matrix-valued operators~$\fh(\hat\h,\la)$ continue to look like
matrix-valued {\it functions\/} in the noncommutative setup.

While formulating Riemann-Hilbert problems for (commutative or
noncommutative) {\it multi-\/}monopoles is straightforward, solving them
is another story. We succeeded here for a certain class of Wu-Yang
multi-monopoles but must otherwise leave this task for future research.

\bigskip

\noindent
{\large{\bf Acknowledgements}}

\medskip
\noindent
O.L. thanks the YITP of SUNY at Stony Brook for hospitality and
in particular M.~Ro\v cek for discussions.
A.D.P. is grateful to N.~Nekrasov for inspiring discussions and
to IHES for hospitality at an intermediate stage of the project.
This work was partially supported by grant Le~838/7 within
the framework of the DFG priority program (SPP 1096) in string theory.
Further support at an early stage came from
a sabbatical research grant of the Volkswagen-Stiftung.

\bigskip

\begin{appendix} 
\section{Appendix: Self-dual gauge fields in the twistor approach} 
 
\noindent 
{\bf SDYM equations.} 
It is well known (see e.g. \cite{Manton, Actor, Atiyah1}) 
that the Bogomolny equations (\ref{bog.eq.}) are equivalent to the  
reduced self-dual Yang-Mills (SDYM) equations in four Euclidean dimensions. 
Namely, let us consider the Euclidean space $\R^4$ with the metric 
$\de_{\mu\nu}$, a gauge potential $A=A_{\mu}\diff x^\mu$ and the Yang-Mills  
field $F=\diff A +A\wedge A= \frac{1}{2}F_{\mu\nu}\diff x^\mu\wedge\diff x^\n$ 
where $\mu,\nu,\ldots=1,2,3,4$. We assume that the fields $A$ and $F$ take 
values in the Lie algebra $u(n)$.  
 
The SDYM equations have the form 
\begin{equation}\label{sdym1} 
*F\ =\ F \qquad\Longrightarrow\qquad 
\sfrac{1}{2}\ve_{\mu\nu\r\s}F_{\r\s}\ =\ F_{\mu\nu}\ , 
\end{equation} 
where $*$ denotes the Hodge star operator and 
$\ve_{\mu\nu\r\s}$ is the completely antisymmetric tensor in  
$\R^4$, with $\ve_{1234}=1$. If one assumes that the $A_\m$ do not depend  
on $x^4$ and puts $\p :=A_4$ then (\ref{sdym1}) reduces to the  
Bogomolny equations (\ref{bog.eq.}). Moreover, under the same conditions  
the energy density in (\ref{bog}) follows from the pure  
Yang-Mills Lagrangian density 
\begin{equation} 
\Lcal \= - \frac{1}{4}\, \tr\, (F_{\mu\nu}F_{\mu\nu}) 
\end{equation} 
in four Euclidean dimensions. 
 
As reduction of the SDYM equations, the Bogomolny equations are seen to be
integrable. Hence, one may apply powerful twistor methods, 
which have been developed for solving the SDYM equations,  
also to the construction of monopole solutions. Therefore, 
we briefly recall the twistor description of self-dual gauge fields
in this appendix.  
 
\noindent 
{\bf Linear system.} 
Let us introduce the complex coordinates 
\begin{equation} \label{coco} 
y=x^1+\im x^2\  ,\quad z=x^3-\im x^4\ ,\quad  
\bar y=x^1-\im x^2\  ,\quad \bar z=x^3+\im x^4 
\end{equation} 
and put 
\begin{equation} 
A_y=\sfrac{1}{2}(A_1-\im A_2)\ ,\quad  
A_z=\sfrac{1}{2}(A_3+\im A_4)\ ,\quad  
A_{\bar y}=\sfrac{1}{2}(A_1+\im A_2)\ ,\quad 
A_{\bar z}=\sfrac{1}{2}(A_3-\im A_4)\ . 
\end{equation} 
Then the SDYM equations (\ref{sdym1}) read 
\begin{equation}\label{sdym2}  
[D_y, D_z]=0\quad ,\qquad [D_{\bar y},D_{\bar z}]=0\quad ,\qquad  
[D_y, D_{\bar y}]+[D_z, D_{\bar z}]=0\ ,  
\end{equation} 
where $D_\mu :=\pa_\mu +A_\mu$. 
 
These are equations for the connection 1-form $A=A_\m\diff x^\m$ 
on a (trivial) complex vector bundle 
\begin{equation}\label{cvb} 
E=\R^4\times\C^n 
\end{equation} 
over $\R^4$ associated to the principal bundle $P=\R^4\times\mbox{U}(n)$. 
They can be obtained as the compatibility conditions 
of the following linear system of equations~\cite{Ward3, BZ}: 
\begin{equation}\label{ls} 
(D_{\bar y} - \la D_z )\,\psi (x,\la, \lb )=0\ , \quad 
(D_{\bar z} + \la D_y )\,\psi (x,\la, \lb )=0\ , \quad\mbox{and}\quad 
\pa_{\lb} \psi (x,\la, \lb )=0\ , 
\end{equation} 
where the $n\times n$ matrix $\psi$ depends on an extra `spectral parameter' 
$\la$ which lies in the extended complex plane $\C P^1=\C\cup\{\infty\}$. 
 
\noindent 
{\bf Twistor space.}  
In fact, the matrix-valued function $\psi$ in (\ref{ls}) is defined  
on the {\it twistor space\/} $\Zcal =\R^4\times S^2$ for  
the space $\R^4$~\cite{Atiyah2, Atiyah3, Ward-Wells} with the 
canonical projection 
\begin{equation}\label{bundle} 
\pi\ :\ \Zcal\ \to\ \R^4\ . 
\end{equation} 
It can be viewed as a collection of $n$ sections of the pulled-back 
vector bundle $\pi^*E$ over $\Zcal$. Then, as we shall explain shortly,
(\ref{ls}) is interpreted as the holomorphicity of the vector bundle 
$\pi^*E \to \Zcal$. 
 
Recall that the Riemann sphere $S^2\cong\C P^1$ can be covered  
by two coordinate patches $U_+$ and $U_-$ with 
\begin{equation} 
\C P^1=U_+\cup  U_-\quad ,\qquad  
U_+=\C P^1\setminus\{\infty\}\quad ,\qquad 
U_-=\C P^1\setminus\{0\}\ , 
\end{equation} 
and coordinates $\la$ and $\tilde\la$ on $U_+$ and $U_-$, respectively. 
Therefore, also $\Zcal$ can be covered by two coordinate patches, 
\begin{equation} 
\Zcal =\U_+\cup\U_-\quad ,\qquad  
\U_+=\R^4\times U_+\quad ,\qquad  
\U_-=\R^4\times U_-\ , 
\end{equation} 
with coordinates $\{x^\mu , \la , \lb \}$ on $\U_+$ and 
$\{x^\mu , \lt , \bar{\lt}\}$ on $\U_-$. 
One can also introduce complex coordinates 
\begin{equation} 
w_1=y-\la\bar z\ ,\ w_2=z+\la \bar y\ ,\ w_3=\la\qquad\mbox{and}\qquad 
\tilde w_1=\tilde \la y-\bar z\ ,\ \tilde w_2=\tilde \la z+\bar y\ ,\  
\tilde w_3=\tilde \la 
\end{equation} 
on $\U_+$ and $\U_-$, respectively. On the intersection 
$\U_+\cap\U_-\cong\R^4\times\C^*$ these coordinates are related by 
\begin{equation}\label{relco} 
\tilde w_1=\frac{w_1}{w_3}\quad ,\qquad \tilde w_2=\frac{w_2}{w_3} 
\qquad\mbox{and}\qquad \tilde w_3=\frac{1}{w_3}\ . 
\end{equation} 
On the open set $\U_+\cap\U_-$ one may use either set of coordinates, 
and we will use $w_1$, $w_2$ and $w_3$. 
 
\noindent
{\bf Complex and real structures on $\Zcal$.} 
It follows from (\ref{relco}) that, as a complex manifold, $\Zcal$ is 
not a direct product $\C^2\times\C P^1$ but  is a nontrivial rank 
two holomorphic vector bundle over $\C P^1$ with a holomorphic
projection
\begin{equation}\label{Zcal} 
p \ : \quad \Zcal = \Ocal(1)\otimes\C^2 = 
\Ocal(1)\oplus\Ocal(1)\ \to\ \C P^1\ , 
\end{equation} 
where $\Ocal(k)$ is the holomorphic line bundle over $\C P^1$ 
with first Chern class $c_1(\Ocal(k))=k$. 
Real holomorphic sections of the bundle (\ref{Zcal}) are the projective lines
\begin{equation}\label{sections} 
\cp\ =\ \begin{cases} 
\la\in U_+,&w_1=y-\la{\zb} ,\quad  w_2=z+\la{\yb} \\ 
\la\in U_-,&\tilde w_1=\tilde\la y-{\zb} , \quad \tilde w_2=\tilde\la z + {\yb}
\end{cases} 
\end{equation} 
parametrized by points $x\in\R^4$. These sections are real in the sense 
that they are invariant w.r.t. an antiholomorphic involution ({\it real  
structure})  $\tau : \ \Zcal\to\Zcal$ defined by 
\begin{equation}\label{realstr} 
\tau (y, \yb, z, \zb, \la) \= (\yb, y, \zb, z, -\lb^{-1})\ . 
\end{equation} 
It is easily seen from (\ref{realstr}) that $\tau$ has no fixed points 
in $\Zcal$ but does leave the projective lines $\cp\hookrightarrow\Zcal$ 
invariant. These Riemann spheres $\cp$ are parametrized by $(x^\m)=(y,\yb, 
z, \zb) \in\R^4$. Note that on matrix-valued functions $\vp (x,\la)$ the  
involution $\tau$ acts as follows~\cite{Manin}: 
\begin{equation} 
\tau \ :\quad \vp (x,\la)\ \mapsto\ (\tau \vp) (x,\la)
\ :=\ [\vp (x, -\lb^{-1})]^\+\ . 
\end{equation} 
 
\noindent 
{\bf Pulled-back bundle $\pi^*E$.} 
Consider the complex vector bundle $E$ 
introduced in (\ref{cvb}) and its pullback $\pi^*E$ to $\Zcal$. Note that 
the fibres of the bundle (\ref{bundle}) over points $x\in\R^4$ are the 
two-spheres (\ref{sections}), i.e. $\pi^{-1}(x)=\cp$. By definition, the 
pulled-back bundle $\pi^*E$ is trivial on the fibres $\cp$. For the 
pulled-back connection $\tilde D:=\pi^*D=D+\diff\la\pa_\la + 
\diff{\lb}\pa_{\lb}$ one can introduce the  $(0,1)$ part $\tilde D^{(0,1)}:=
\pi^*D^{(0,1)}=\bar{\pa} +(\pi^*A)^{(0,1)}$ w.r.t. the complex structure 
on $\Zcal$. Its components along antiholomorphic vector fields 
on $\U_+$, 
\begin{equation}\label{vf} 
\bar V_{\bar 1}\=\pa_{\bar y}-\la\pa_z\ , \qquad 
\bar V_{\bar 2}\=\pa_{\bar z}+\la\pa_y\ , \qquad  
\bar V_{\bar 3}\=\pa_{\lb} \ ,
\end{equation} 
have the form (see e.g.~\cite{Popov:1999pc} and references therein) 
\begin{equation} 
\tilde D^{(0,1)}_{\bar 1}=\ \bar V_{\bar 1} + A_{\yb}-\la A_z\ , \qquad 
\tilde D^{(0,1)}_{\bar 2}=\ \bar V_{\bar 2} + A_{\zb}+\la A_y\ , \qquad 
\tilde D^{(0,1)}_{\bar 3}=\ \bar V_{\bar 3}\ , 
\end{equation} 
which coincide with those in (\ref{ls}). 
 
\noindent
{\bf Holomorphic sections of $\pi^*E$.} 
Let us consider local holomorphic sections $s$ of the bundle 
$\pi^*E\to\Zcal$, which are defined by the equation $\tilde D^{(0,1)}s=0$ or, 
in components, 
\begin{equation}\label{ls2} 
\tilde D^{(0,1)}_{\bar a}s=0 \qquad\textrm{for}\quad \bar a=1,2,3\ . 
\end{equation} 
The bundle $E':=\pi^*E$ is called holomorphic iff these equations are 
compatible in the sense that $(\tilde D^{(0,1)})^2=0\ \Leftrightarrow \  
[\tilde D^{(0,1)}_{\bar a},\tilde D^{(0,1)}_{\bar b}]=0\ $ for 
$\ \bar a, \bar b,\ldots =1,2,3\ $. 
These compatibility equations coincide with the SDYM equations  
(\ref{sdym2}). So, if the gauge field is self-dual then there 
exist local solutions $s_\pm$ on $\U_\pm$ 
with $s_+=s_-$ on the intersection $\U_+\cap\U_-$. Note that one can always  
represent $s_\pm$ in the form $s_\pm=\psi_\pm\chi_\pm$, where $\psi_\pm$ are 
$n\times n$ matrix-valued nonsingular functions on $\U_\pm$ satisfying 
$\tilde D^{(0,1)}\psi_\pm =0$, and $\chi_\pm\in\C^n$ are 
vector functions satisfying $\bar\pa\chi_\pm =0$. 

\noindent
{\bf Holomorphic transition functions.}
The vector functions $\chi_\pm\in\C^n$ are holomorphic on 
$\U_\pm$, i.e. they are only functions of $w_a$ and 
$\tilde w_a$, respectively. Furthermore, on $\U_+\cap\U_-$ they
are related via 
\begin{equation} 
\chi_+\=f_{+-}\,\chi_- 
\end{equation} 
with 
\begin{equation}\label{trfun} 
f_{+-}\=\psi_+^{-1}\,\psi_- 
\end{equation} 
which implies holomorphicity of $f_{+-}$ on $\U_+\cap\U_-$, i.e. 
\begin{equation}\label{hlm} 
(\pa_{\bar y}-\la\pa_z)f_{+-}=0 \ ,\qquad 
(\pa_{\bar z}+\la\pa_y)f_{+-}=0\qquad\mbox{and}\qquad  
\pa_{\lb} f_{+-}=0\ . 
\end{equation} 
One may restrict this function to the manifold $\Zcal_0=\R^4\times S^1\subset 
\U_+\cap\U_-\subset\Zcal$ on which it will be real-analytic. Any such function 
defines a holomorphic bundle $E'$ over $\Zcal$. Namely, $f_{+-}$ can be 
identified  
with a transition function in the holomorphic bundle $\pi^*E$ over $\Zcal$, 
and the pair of functions $\psi_\pm$ in (\ref{trfun}) defines a smooth 
trivialization of this bundle.  

\noindent
{\bf Reality conditions.}
The reality of the gauge fields is an important issue~\cite{Atiyah2, Atiyah3}. 
The antihermiticity conditions $A_\mu^\+ = -A_\mu$ for the components of the 
gauge potential can be satisfied by imposing the following 
`reality' conditions for the matrices   $f_{+-}$ and $\psi_{\pm}$: 
\begin{equation}\label{cond2}  
f_{+-}^\+ (x, -{\bar\la}^{-1})\ =\ f_{+-} (x,\la)\ , 
\end{equation} 
\begin{equation}\label{cond1}  
\psi_+^\+(x, - {\bar\la}^{-1})\ =\ \psi_-^{-1}(x,\la)\ . 
\end{equation} 
It is easy to see that the gauge transformations 
\begin{equation} 
A_\mu\quad \mapsto\quad A_\mu^g\ =\ g^{-1}A_\mu\, g\ +\ g^{-1}\pa_\mu\, g 
\end{equation} 
are induced (via (\ref{ls})) by the transformations 
\begin{equation} 
\psi_+\quad \mapsto\quad \psi_+^g\ =\ g^{-1}\psi_+\qquad\textrm{and}\qquad 
\psi_-\quad \mapsto\quad \psi_-^g\ =\ g^{-1}\psi_-\ , 
\end{equation} 
where $g{=}g(x)$ is an arbitrary U$(n)$-valued function on~$\R^4$. 
The transition function $f_{+-}=\psi_+^{-1}\psi_-$ is obviously invariant  
under these transformations. Hence, we have shown that one can associate 
a transition function $f_{+-}$ in the holomorphic vector bundle $\pi^*E$ 
over the twistor space $\Zcal$ to a gauge equivalence class $[A]$ of 
solutions to the SDYM equations on~$\R^4$. In other words, we have 
described a map from complex vector bundles $E$ over~$\R^4$ with self-dual 
connections (modulo gauge transformations) into topologically trivial  
holomorphic vector bundles $E'=\pi^*E$ over the twistor space $\Zcal\cong 
\R^4\times S^2$. 
 
\noindent 
{\bf Splitting of transition functions.}  
Consider now the converse situation. Let $E'$ be a 
topologically trivial holomorphic bundle 
over $\Zcal$ with a real-analytic transition function $f_{+-}$  
on $\Zcal_0=\R^4\times S^1\subset \Zcal$. Suppose that a restriction  
of $E'$ to any projective line $\cp\hookrightarrow\Zcal$ (a section 
(\ref{sections}) of the bundle (\ref{Zcal})) is holomorphically  
trivial. Then for each fixed $x\in\R^4$ one can find matrix-valued  
functions $\psi_{\pm}(x,\la)$ such that $f_{+-}(x,\la)$ as a function of  
$\la\in S^1\subset\C P^1$ can be split as $f_{+-}(x,\la)= 
\psi_+^{-1}(x,\la)\psi_-(x,\la)$, where $\psi_+$ and $\psi_-$ are  
boundary values of holomorphic functions on (subsets of) $U_+\subset\C P^1$ 
and  $U_-\subset\C P^1$, respectively. For a fixed point $x\in\R^4$, 
the task to split  a matrix-valued real-analytic function 
$f_{+-}(x,\la)=f_{+-}(y{-}\la{\zb},z{+}\la{\yb},\la)$ on 
\begin{equation}\label{s1x}
S^1_x\=\{\la\in\C P^1\ :\ |\la |=1,\ w_1=y-\la\zb ,\ w_2=z+\la\yb\}
\end{equation} 
defines a parametric Riemann-Hilbert problem on 
${\cp}{\hookrightarrow}{\Zcal}$. If its solution exists 
for given $x\in\R^4$ then one can prove that solutions exist also on an open 
neighbourhood $U$ of $x$ in $\R^4$~\cite{Ward-Wells}. 
 
{} From the holomorphicity of $f_{+-}$ it follows that $f_{+-}(x,\la)= 
f_{+-}(y{-}\la{\zb}, z{+}\la{\yb},\la)$ satisfies (\ref{hlm}). After 
splitting $f_{+-}(x,\la)=\psi_+^{-1}(x,\la)\psi_-(x,\la)$ we obtain 
\begin{equation}\label{hol} 
\psi_{+}(\pa_{\bar y}-\la\pa_z)\psi_{+}^{-1} = 
\psi_{-} (\pa_{\bar y}-\la\pa_z)\psi_-^{-1} \quad\ \textrm{and}\quad\ 
\psi_{+}(\pa_{\bar z}+\la\pa_y)\psi_{+}^{-1} = 
\psi_{-} (\pa_{\bar z}+\la\pa_y)\psi_-^{-1}\ . 
\end{equation} 
We expand $\psi_+$ and $\psi_-$ into power series in $\la$ and $\la^{-1}$,  
respectively. 
Upon substituting into~(\ref{hol}) one easily sees that both sides  
of~(\ref{hol}) must be linear in $\la$, and one can introduce 
Lie-algebra valued fields $A_\mu$ by 
\begin{equation}\label{A1} 
A_{\bar y}-\la A_z\ :=\ \psi_{\pm}(\pa_{\bar y}-\la\pa_z)\psi_{\pm}^{-1} 
\qquad\textrm{and}\qquad 
A_{\bar z}+\la A_y\ :=\ \psi_{\pm}(\pa_{\bar z}+\la\pa_y)\psi_{\pm}^{-1} \ . 
\end{equation} 
Hence, the gauge potential components may be calculated from 
\begin{equation}\label{A2} 
A_{\bar y}\ =\ \psi_{+}\pa_{\bar y}\psi_{+}^{-1}|_{\la =0}\ =\ 
-A_y^\+ \qquad\textrm{and}\qquad 
A_{\bar z}\ =\ \psi_{+}\pa_{\bar z}\psi_{+}^{-1}|_{\la =0}\ =\ 
-A_z^\+ 
\end{equation}
if $\psi_\pm$ satisfy the reality condition (\ref{cond1}).
By construction, the components $A_\mu$ of the gauge potential $A$  
defined above satisfy the SDYM equations. 
Note that the gauge potential $A$ is inert under the transformations 
\begin{equation} \label{inert} 
\psi_+\ \mapsto\ \psi_+\,h_+^{-1}\ , \quad 
\psi_-\ \mapsto\ \psi_-\,h_-^{-1}\ , \quad 
f_{+-}\ \mapsto\ h_+f_{+-}h_-^{-1}\ , 
\end{equation} 
where $h_+{=}h_+(w_1,w_2,\la)$ and 
$h_-{=}h_-(\tilde w_1,\tilde w_2,\tilde\la)$  
are arbitrary matrix-valued holomorphic functions on $\U_+$ and $\U_-$,  
respectively. Two transition functions are said to be equivalent if they are 
related by the transformation (\ref{inert}). Thus, it defines an  
equivalence class $[E']$ of holomorphic bundles $E'$. 
 
\noindent 
{\bf Noncommutative twistor correspondence.} 
To sum up, we have described a  
one-to-one correspondence between pairs $(E,[D])$ and $([E'], \bar{\pa} )$, 
where $E$ is a complex vector bundle over $\R^4$, $[D]$ are gauge equivalence  
classes of self-dual connections on $E$, and $[E']$ are equivalence classes 
of holomorphic vector bundles $E'$ over the twistor space $\Zcal$ for $\R^4$ 
which are holomorphically trivial on each real projective line $\cp$ in $\Zcal$,
$x\in U\subset\R^4$. This so-called {\it twistor correspondence\/} can be 
described by the following diagram, 
\begin{equation}\label{d1} 
\begin{CD} 
E'@>>>{\Zcal} @>p>>{\C P^1} \\ 
@.   @V{\pi}VV    \\ 
E@>>>{\R^4}@. 
\end{CD} 
\end{equation} 
 
Note that instead of the complex vector bundle $E$ and holomorphic vector 
bundle $E'$ one can consider the space $\Ecal = \Acal (\R^4)\otimes\C^n$ 
of (smooth or real-analytic) sections of $E$ and the sheaf $\Ecal '$ of 
holomorphic sections of $E'$, respectively. Then the twistor correspondence 
can be reformulated as a correspondence between pairs $(\Ecal,[D])$
and $(\Ecal',\bar{\pa})$. 
To obtain a noncommutative deformation of the twistor correspondence 
one should simply introduce sheaves $\Ecal_{\th}$ and $\Ecal_{\th}'$ 
(parametrized by some constant matrix $\th$) which are deformations 
of $\Ecal$ and $\Ecal '$, respectively~\cite{Kapustin}. 
 
\noindent
{\bf Deformed sheaves of functions.} 
For a realization of the above-mentioned deformations, consider 
the tensor algebra $T(\R^4)$ over $\R^4$ and a two-sided ideal $I$ 
in $T(\R^4)$  generated by elements of the form 
\begin{equation} 
\hat{r}^{\m\n}\ :=\ \xh^\m\xh^\n - \xh^\n\xh^\m - \im\th^{\m\n}\ , 
\end{equation} 
where $\th = (\th^{\m\n})$ is an antisymmetric constant matrix. 
Then, define the algebra $\Acal_{\th} (\R^4)$ as an algebra over $\C$ 
generated by $\xh^\m$ with the relations 
\begin{equation}\label{comxh} 
[\xh^\m , \xh^\n ] \= \im\th^{\m\n}\ , 
\end{equation} 
i.e. as the quotient algebra 
\begin{equation} 
\Acal_{\th} (\R^4) \= T (\R^4) / I 
\end{equation} 
usually called the Weyl algebra. Deviating from the standard notation, we denote
by $\R^4_{\th}$ the real vector space spanned by $\xh^1, \xh^2, \xh^3, \xh^4$.
Clearly, $\R^4_{\th}\subset\Acal_{\th} (\R^4)$. Analogously, we denote by  
$\C^2_\th(\la)$ and $\C^2_\th(\lt)$ the complex vector spaces spanned by 
$\wh_1=\yh{-}\la\hat{\zb}$ and $\wh_2=\zh{+}\la\hat{\yb}$ 
with $\la\in U_+\subset \C P^1$, and by 
$\hat{\tw}_1=\lt\yh{-}\hat{\zb}$ and $\hat{\tw}_2=\lt\zh{+}\hat{\yb}$ 
with $\lt\in U_-\subset\C P^1$, respectively. As in the commutative case,  
$\wh_{1,2}$ and  $\hat{\tw}_{1,2}$ are related on $U_+\cap U_-$ by 
$\wh_{1,2}=\la\hat{\tw}_{1,2}$, i.e. they are operator-valued sections over 
$U_+$ and $U_-$ of the holomorphic bundle $\Ocal (1)\otimes\C^2$. Their 
commutation relations follow from (\ref{comxh}). They generate algebras 
$\Ocal_{\th}(\C^2 (\la))\supset \C_{\th}^2 (\la)$ and $\Ocal_{\th} (\C^2  
(\lt ))\supset \C_{\th}^2 (\lt )$ of holomorphic functions of $\wh_1, \wh_2$  
and $\hat{\tw}_1, \hat{\tw}_2$, respectively. These algebras are 
subalgebras in the algebras  $\Ocal_{\th}(\U_{\pm})$ of local sections 
of the holomorphic sheaf $\Ocal_{\th}$ over $\U_{\pm}\subset\Zcal$ 
generated by $\wh_a$ and $\hat{\tw}_a$. Note that we deform the 
spaces ${\cal E}'(\U_{\pm})$ of local holomorphic sections of $E'$ over 
$\U_+$ and $\U_-$ instead of considering 
global smooth (or real-analytic) sections and a projector encoding transition  
functions $f_{+-}$ in $E'$ via the Serre-Swan theorem~\cite{Swan}. The reason 
is that the Riemann sphere $\C P^1$ in (\ref{d1}) is not deformed, and we 
consider operator-valued functions on $U_\pm\subset\C P^1$ glued on the 
overlap $U_+\cap U_-$ by operator transition functions $\fh_{+-}(\xh ,\la)$ 
when discussing Riemann-Hilbert problems and Birkhoff factorizations. 

\noindent
{\bf Deformed twistor space.} 
One can introduce the deformed twistor space as 
$\Zcal_{\th}':= \Acal_{\th} (\R^4) \times\C P^1$ with a double fibration 
\begin{equation}\label{d2} 
\begin{CD} 
\Acal_{\th} (\R^4)\times\C P^1 @>p'>>{\C P^1} \\ 
@V{\pi '}VV    \\ 
\Acal_{\th} (\R^4)@. 
\end{CD} 
\end{equation} 
where the fibres of the holomorphic projection $p'$ are the algebras  
$\Ocal_{\th}(\C^2 (\la))= p'^{-1}(\la)$ for $\la\in\C P^1$. Instead of 
(\ref{d2}) one may also consider the space 
$\Zcal_{\th} :=\R^4_{\th}\times\C P^1$ with two projections, 
\begin{equation}\label{d3} 
\begin{CD} 
\R^4_{\th}\times\C P^1 @>p>>{\C P^1} \\ 
@V{\pi}VV    \\ 
\R^4_\th @. 
\end{CD} 
\end{equation} 
for a tighter connection with the commutative case. In (\ref{d3}) the fibres 
over $\C P^1$ are the vector spaces $\C^2_\th (\la')= p^{-1}(\la')$ for 
$\la'\in {\C P^1}$. 

\noindent
{\bf Operator Riemann-Hilbert problems.}
We will not discuss the noncommutative twistor correspondence in detail since 
it mainly repeats the commutative one with $x^\m\mapsto\xh^\m$. In basic terms,
there is a correspondence between gauge equivalence classes of operators 
$\hat{A}_\m (\xh)$ solving the noncommutative SDYM equations and 
holomorphic equivalence classes of (operator-valued) transition functions 
$\fh_{+-}(\xh,\la)=\fh_{+-}(\yh{-}\la\hat{\zb},\zh{+}\la\hat{\yb},\la)$ 
in the complex vector bundle over $\C P^1$ with fibres 
$\Ocal_{\th}(\C^2 (\la))\otimes\C^n$ at $\la\in\C P^1$, 
such that $\fh_{+-}(\xh,\la)=\hat\psi_+^{-1}(\xh,\la) \hat\psi_-(\xh,\la)$ 
on $S^1\subset\C P^1$ and one has
\begin{equation}\label{hatA} 
\hat{A}_{\bar y}-\la \hat{A}_z\= 
\hat\psi_{\pm}(\pa_{\bar y}-\la\pa_z)\hat\psi_{\pm}^{-1} 
\qquad\textrm{and}\qquad 
\hat{A}_{\bar z}+\la \hat{A}_y\= 
\hat\psi_{\pm}(\pa_{\bar z}+\la\pa_y)\hat\psi_{\pm}^{-1} \ . 
\end{equation} 
Here, $\hat\psi_+(\xh,\la)$ and $\hat\psi_-(\xh,\la)$ are boundary values of 
a holomorphic function (in $\la$) on (subsets of) $U_+\subset\C P^1$ and 
$U_-\subset\C P^1$, respectively. They solve the operator Riemann-Hilbert
problem of splitting $\fh_{+-}(\xh,\la)$ on $S^1 =\{\la\in\C P^1 : |\la |=1\}$.
{}From (\ref{hatA}) one can either find $\hat\psi_\pm(\xh,\la)$ and therefore 
$\fh_{+-}(\xh,\la)$ if the $\hat{A}_{\m}(\xh)$ are given or, conversely, find
the $\hat{A}_{\m}(\xh)$ if $\fh_{+-}(\xh,\la)$ and its Birkhoff decomposition 
are known. 
 
\bigskip

\section{Appendix: Monopoles in the twistor approach}

\noindent
{\bf Translational invariance.}
As was briefly mentioned in appendix A, the Bogomolny equations in $\R^3$
can be obtained from the self-dual Yang-Mills (SDYM) equations in $\R^4$ by 
imposing on the gauge potential $\{A_\m , \m =1,\ldots,4\}$ a condition of 
invariance w.r.t. translation along $x^4$-axis 
and by putting $\p :=A_4$. Here we are going to consider
the twistor description of monopoles which follows from the description of
self-dual gauge fields discussed in appendix A.

The translation generated by the vector field $T=\pa /\pa x^4$ is an isometry 
of $\R^4$ and yields a free twistor space action of the additive group $\R$ 
which is the real part of the holomorphic action of the complex number $\C$. 
This means that smooth 
$T$-invariant functions on  $\Zcal$ can be considered as `free' functions 
on the manifold
\begin{equation}
Y=\R^3\times S^2\ .
\end{equation}
At the same time, for holomorphic functions $f$ on $\Zcal$ we have
\begin{equation}
Tf\=\pa_4 f(w_a)\=\pa_4 w_a\frac{\pa}{\pa w_a} f
\= -\im (\pa_{w_2}+\la\pa_{w_1}) f(w_a)\ =:\ T'f \ ,
\end{equation}
where
\begin{equation}\label{t'}
T' \= -\im (\pa_{w_2}+\la\pa_{w_1})\=-\im (\pa_{\tw_1}+\lt\pa_{\tw_2})\ .
\end{equation}
In other words, $T=T'+{\overline T}'$, 
where the bar denotes complex conjugation.
The first expression in (\ref{t'}) is valid on $\U_+$
and the second one on $\U_-$ (see appendix A for the definition of $w_a, \tw_a,
\U_\pm$ and $\Zcal$). So, $T'$-invariant holomorphic functions on $\Zcal$
can be considered as `free' holomorphic functions on a reduced twistor space
$\cal T$ (called minitwistor space~\cite{Hitchin, Hurtubise}) obtained as the 
quotient space of $\Zcal$ by the action of the complex abelian symmetry group
generated by $T'$.

\noindent
{\bf Minitwistor space $\cal T$.}
It is not difficult to see that $\cal T$ is covered by two coordinate patches
$V_+$ and $V_-$ with coordinates 
$(\h=w_1{-}\la w_2,\la)$ and $(\tilde\h=\tw_2{-}\lt\tw_1,\lt)$, respectively.
These coordinates are constant along the $T'$-orbits in $\Zcal$, 
and on the overlap $V_+\cap V_-$ we have
\begin{equation}\label{hl}
\h =\la^2\tilde\h \qquad\textrm{and}\qquad \la = \lt^{-1}
\end{equation}
which coincides with the transformation of coordinates on the total space of
the holomorphic line bundle $\Ocal (2)\to \C P^1$ with $c_1(\Ocal (2))=2$. 
In (\ref{hl}), $\la$ and $\lt$ are coordinates on $\C P^1$ (see appendix A), 
while $\h$ and $\tilde\h$ are coordinates in the fibres over $U_+$ and $U_-$, 
respectively, where $U_+\cup U_-=\C P^1$. So, as the minitwistor space we obtain
the total space $\cal T$ of the bundle
\begin{equation}\label{Tcal}
p:\quad {\cal T}=\Ocal (2) \cong T\C P^1\ \to \ \C P^1\ ,
\end{equation}
where $T\C P^1$ denotes the holomorphic tangent bundle of $\C P^1$.
The space $\cal T$ is  part of the double fibration
\begin{equation}\label{d4}
\begin{CD}
Y=\R^3\times S^2@>q>>{\cal T}\\
@V{\pi}VV\\
\R^3
\end{CD} 
\end{equation}
where $\pi$ projects onto $\R^3$ and 
$q(x,\la)=\{\h=y{-}2\la x^3{-}\la^2\yb ,\la\}$.

\noindent
{\bf Real structure on $\cal T$.}
The real structure $\tau$ on $\Zcal$ (see appendix A) induces a real structure
(antiholomorphic involution, $\tau^2=1$) on $\cal T$ acting on (local) 
coordinates as follows,
\begin{equation}\label{rstr}
\tau (\h ,\la) \= (-\bar{\h}/\lb^2, -1/\lb)\ .
\end{equation}
{}From this definition it is obvious that $\tau$ has no fixed points on
$\cal T$ but does leave invariant the projective lines
\begin{equation}\label{prlns}
\cp\ =\ 
\begin{cases}
\la\in U_+\ , & \h = y-2\la x^3 -\la^2\yb  \\
\lt\in U_-\ , & \tilde\h = \lt^2y-2\lt x^3-\yb\end{cases}
\end{equation}
which are real holomorphic sections of the bundle (\ref{Tcal})
parametrized by $x\in\R^3$. Note that the diagram (\ref{d4})
describes a one-to-one correspondence between points $x\in\R^3$ and
projective lines (\ref{prlns}) in $\cal T$. Conversely, points of
$\cal T$ correspond to oriented lines in $\R^3$~\cite{Hitchin, Hurtubise}.

\noindent
{\bf Minitwistor correspondence.}
Recall again that the Bogomolny equations (\ref{bog.eq.}) are a reduction to
$\R^3$ of the SDYM equations on $\R^4$ which arise as compatibility conditions
of the linear system (\ref{ls}) from the appendix A. One can 
discard the ignorable coordinate $x^4$ in the gauge potential
and obtain a reduced linear
system for which the compatibility conditions are equivalent to the
Bogomolny equations. Therefore, a correspondence between solutions $(A_a, \p )$ 
of the Bogomolny equations and holomorphic vector bundles $E'$ over the 
minitwistor space $\cal T$ directly follows from the twistor correspondence
described in appendix~A. Simply, $(A_a, \p{=}A_4)$ and $f_{+-}$ should not
depend on $x^4$, but $x^4$~dependence is allowed for $\psi_\pm (x,\la)$
especially if one abandons the reality condition (\ref{cond1}) for them.
Hence, there is a one-to-one correspondence between pairs $(E, [A_a,\p ])$
and $([E'], \bar{\pa} )$, where $E$ is a complex vector bundle over $\R^3$, 
$[A_a,\p ]$ are gauge equivalence classes of solutions to the Bogomolny 
equations, and $[E']$ are equivalence classes of holomorphic vector
bundles over the minitwistor space $\cal T$ which are holomorphically
trivial on each real projective line $\cp\hookrightarrow\cal T$, 
as defined by (\ref{prlns}).

{\bf Riemann-Hilbert problems.}
Recall that any holomorphic vector bundle $E'$ over $\cal T$ is defined by
a transition function $f_{+-}(\h , \la)$ on $V_+\cap V_-$ whose restriction 
$f_{+-}(y{-}2\la x^3{-}\la^2\yb ,\la)$ to $\cp\hookrightarrow\cal T$ defines
parametric Riemann-Hilbert problems. Therefore, the above-mentioned
correspondence implies a correspondence between solutions  $(A_a, \p )$
of the Bogomolny equations and solutions $\psi_\pm (x,\la)$ of the
Riemann-Hilbert problems on $\cp\hookrightarrow\cal T$, 
where $x\in U\subset\R^3$.
Here, $\psi_\pm (x,\la)$ are the result of a Birkhoff decomposition
$f_{+-}(x,\la)=\psi_+^{-1}(x,\la)\Lambda(x,\la)\psi_-(x,\la)$ with $\Lambda=1$.

Note that $f_{+-}(\h,\la)\equiv f_{+-}(w_1{-}\la w_2,\la)$ is annihilated 
by the vector fields (\ref{vf}).  
Therefore, from the (generalized) Liouville theorem it follows that
$$
\psi_{+}(\pa_{\bar y}-\la\pa_z)\psi_+^{-1} \=
\psi_{-}(\pa_{\bar y}-\la\pa_z)\psi_-^{-1} \=
\sfrac{1}{2}(A_1 +\im A_2)-\sfrac{\la}{2}(A_3 +\im \p )\ ,
$$
\begin{equation}\label{AA}
\psi_{+}(\pa_{\bar z}+\la\pa_y)\psi_+^{-1} \=
\psi_{-}(\pa_{\bar z}+\la\pa_y)\psi_-^{-1} \=
\sfrac{1}{2}(A_3 - \im \p ) + \sfrac{\la}{2}(A_1 - \im A_2)\ ,
\end{equation}
where the fields $(A_a, \phi )$ do not depend on $\la$.
Here we keep $\pa_4$ since, as mentioned above, $\psi_\pm (x,\la)$
may depend on $x^4$ even if $f_{+-}$ and $(A_a,\p )$ do not.
So, if we know $f_{+-}$ and its Birkhoff decomposition $f_{+-}(x,\la)=
\psi_+^{-1}(x,\la)\psi_-(x,\la)$, then $(A_a,\p )$ can be calculated from 
$$
\sfrac{1}{2}(A_1+\im A_2) =
\psi_+(x,\la)\pa_{\bar y}\psi_+^{-1}(x,\la)|_{\la =0}\ ,
\quad
\sfrac{1}{2}(A_3-\im \p ) =
\psi_+(x,\la)\pa_{\bar z}\psi_+^{-1}(x,\la)|_{\la =0}\ ,
\phantom{XXXl}
$$
\begin{equation}\label{AAAA}
\sfrac{1}{2}(A_1 - \im A_2)=\psi_-(x,\la)\pa_{ y}
\psi_-^{-1}(x,\la)|_{\la =\infty}\ ,\quad
\sfrac{1}{2}(A_3 +\im \p )=\psi_-(x,\la)\pa_{z}
\psi_-^{-1}(x,\la)|_{\la =\infty}\ .
\end{equation}
By construction, the fields (\ref{AAAA}) satisfy the Bogomolny equations. 
Furthermore, $(A_a,\p)$ are real (antihermitean) if $\psi_\pm (x,\la)$ satisfy
the reality conditions (\ref{cond1}).

\noindent
{\bf Deformed sheaves of functions.} 
For the noncommutative deformation of the minitwistor correspondence
we consider the algebra $\Acal_{\th}(\R^3)$ introduced in section 2.3,
\begin{equation}
\Acal_{\th} (\R^3) = T (\R^3) / {\cal I}\ ,
\end{equation}
where $T(\R^3)$ is the tensor algebra of $\R^3$, and an ideal $\cal I$
is generated by elements of the form 
\begin{equation}
\xh^1\xh^3-\xh^3\xh^1 \ ,\qquad
\xh^2\xh^3-\xh^3\xh^2 \qquad\textrm{and}\qquad
\xh^1\xh^2-\xh^2\xh^1-\im\th \qquad\textrm{with}\quad 
\th\ge 0\ . 
\end{equation}
Following the logic of appendix A, we denote by $\R^3_\th$ the
real vector space spanned on $\xh^1, \xh^2$ and $\xh^3$. Obviously,
$\R^3_\th\subset\Acal_{\th}(\R^3)$, and $\xh^3$ is the operator of 
multiplication with $x^3$, which commutes with $\xh^1$ and $\xh^2$. 
We also denote by $\C_\th (\la')$
a one-parameter space of operators
\begin{equation}\label{oprtrs}
{\hat\h} (\la) = {\yh}-2\la x^3-\la^2{\hat{\yb}}
\quad \mbox{on}\quad U_+
\qquad\mbox{and}\qquad
{\hat{\tilde\eta}} ({\lt} ) 
= {\lt}^2{\yh}-2{\lt} x^3-\hat{\yb}
\quad \mbox{on}\quad U_-\ .
\end{equation}
On the intersection $U_+\cap U_-$ these operators are related by 
${\hat\h} = \la^2{\hat{\tilde\h}}$ (cf.~(\ref{hl})) with $\la =\lt^{-1}$,
i.e. they are operator-valued sections of the bundle
\begin{equation}\label{calT}
{\cal T}_{\th}\ \to\ \C P^1
\end{equation}
with $\C_\th(\la')$ as fibres over $\la'\in \C P^1$. The spaces $\C_\th(\la')$
generate a one parameter family of algebras $\Ocal_{\th}(\C (\la'))\supset
\C_\th(\la')$ of holomorphic operator-valued functions on the fibres 
of~(\ref{calT}).

\noindent
{\bf Deformed minitwistor space.}
Recall that the real holomorphic sections (\ref{prlns}) of the 
bundle~(\ref{Tcal}) are parametrized by the coordinates $x^a$ on~$\R^3$. 
Substituting operators $\xh^a$ instead of coordinates $x^a$, 
one obtains sections (\ref{oprtrs}) of the bundle (\ref{calT}) which may be 
considered as a noncommutative minitwistor space. Thus, the deformation of 
the minitwistor space $\cal T$ simply means the usage of operators 
$\hat\h$ and $\hat{\tilde\eta}$ instead of coordinates $\h$ and $\tilde\h$ 
on fibres of the bundle (\ref{calT}).
However, this substitution does not change the mutual commutativity of the
`coordinates' $\hat\h$, $\la$ and $\hat{\tilde\eta}$, $\lt$ on $\cal T_{\th}$. 
Note that instead of $\cal T_{\th}$ one may also consider for the deformed
minitwistor space the bundle
\begin{equation}\label{clT}
{\cal T}_{\th}'\ \to\ \C P^1
\end{equation}
with algebras $\Ocal_{\th}(\C (\la'))\supset\C_\th (\la')$ as fibres at 
$\la'\in \C P^1$ (cf.~discussion in appendix A).
In other words, we deform only the fibres in our vector bundles over $\C P^1$ 
since the base manifold $\C P^1$ is unchanged.
Hence, as a generalization of the diagram (\ref{d4}) one may use the double
fibrations
\begin{equation}\label{d5}
\begin{CD}
\Acal_{\th} (\R^3)\times\C P^1 @>>>{\cal T}_{\th}' \\
@VVV    \\
\Acal_{\th} (\R^3)@.
\end{CD}
\qquad\textrm{and}\qquad
\begin{CD}
\R^3_{\th}\times\C P^1 @>>>{\cal T}_{\th} \\
@VVV    \\
\R^3_\th @.
\end{CD}
\end{equation}
for a description of the noncommutative minitwistor correspondence.

\noindent
{\bf Operator Riemann-Hilbert problems.}
Finally, a correspondence between solutions 
of the noncommutative Bogomolny equations and 
solutions of the operator Riemann-Hilbert problems is described by
formulae generalizing (\ref{AA}),
$$
\hat\psi_{+}(\xh ,\la)(\pa_{\bar y}-\la\pa_z)\hat\psi_{+}^{-1}(\xh ,\la) =
\hat\psi_{-}(\xh ,\la)(\pa_{\bar y}-\la\pa_z)\hat\psi_-^{-1}(\xh ,\la) =
\sfrac{1}{2}(\hat A_1 +\im \hat A_2)-\sfrac{\la}{2}(\hat A_3 +\im \hat\p )\ ,
\phantom{XXX}
$$
\begin{equation}\label{hatAA}
\hat\psi_{+}(\xh ,\la)(\pa_{\bar z}+\la\pa_y)\hat\psi_{+}^{-1}(\xh ,\la) =
\hat\psi_{-}(\xh ,\la)(\pa_{\bar z}+\la\pa_y)\hat\psi_-^{-1}(\xh ,\la) = 
\sfrac{1}{2}(\hat A_3 - \im \hat\p )+\sfrac{\la}{2}(\hat A_1 -\im \hat A_2)\ .
\end{equation}
The derivation of these lines literally follows that of~(\ref{AA}). 
Just note that now $\fh_{+-}({\yh}{-}2\la x^3{-}\la^2{\hat{\yb}},\la)$
is an operator transition function in a holomorphic vector bundle over
$\C P^1$ with fibres $\Ocal_{\th}(\C(\la'))\otimes\C^n$ at $\la'\in \C P^1$.
So, for a given $\fh_{+-}(\xh,\la)=\hat\psi_+^{-1}(\xh,\la)\hat\psi_-(\xh,\la)$
one can calculate $(\hat A_a,\hat\p )$ from 
$$
\sfrac{1}{2}(\hat A_1 +\im \hat A_2)=\hat\psi_+(\xh ,\la)\pa_{\bar y}
\hat\psi_+^{-1}(\xh ,\la)|_{\la =0}\ ,\quad
\sfrac{1}{2}(\hat A_3 - \im \hat\p )=\hat\psi_+(\xh ,\la)\pa_{\bar z}
\hat\psi_+^{-1}(\xh ,\la)|_{\la =0}\ ,
\phantom{XXX}
$$
\begin{equation}\label{opflds}
\sfrac{1}{2}(\hat A_1 - \im \hat A_2)=\hat\psi_-(\xh ,\lt )\pa_{ y}
\hat\psi_-^{-1}(\xh ,\lt )|_{\lt =0}\ ,\quad
\sfrac{1}{2}(\hat A_3 +\im \hat\p )=\hat\psi_-(\xh ,\lt )\pa_{z}
\hat\psi_-^{-1}(\xh ,\lt )|_{\lt =0}\ .
\end{equation}
By construction, these fields satisfy the noncommutative Bogomolny equations. 

\bigskip

\section{Appendix: General Wu-Yang multi-monopoles}

\noindent
In section~5.2 we have described the Wu-Yang multi-monopole solution 
generated by the transition function~(\ref{fSU2}) with $\r$ 
given by~(\ref{rn}). This transition function belongs to the so-called 
Atiyah-Ward ${\cal A}_1$ ansatz. More general Wu-Yang SU(2) multi-monopoles 
can be obtained from the Atiyah-Ward ${\cal A}_n$ ansatz
with a transition function of the form
\begin{equation}\label{fSU2n}
f_{+-}^{(n)WY}\=
\begin{pmatrix}
\r_n &\la^{-n}\\ -\la^n &0
\end{pmatrix} \ ,
\end{equation}
where $\r_n$ coincides with the function $f_{+-}^{(n)D}=(f_{-+}^{(n)D})^{-1}$
(see~(\ref{trfun2})) generating the Dirac multi-monopole,
\begin{equation}
\r_n \= \prod_{k=1}^n \frac{\la}{\h-h_k} \= f_{+-}^{(n)D} \ .
\end{equation}
Note that the splitting $f_{+-}^{(n)WY}=(\psi_+^{(n)})^{-1}\psi_-^{(n)}$
of the ${\cal A}_n$ matrix~(\ref{fSU2n}) is a more complicated task than
for the ${\cal A}_1$ case considered in section~5, and we will not consider 
it here.

Via splitting of the matrix~(\ref{fSU2n}) one can obtain the generic
Wu-Yang SU(2) multi-monopole configuration which is precisely equivalent
to the Dirac multi-monopole~(\ref{Ap}). This statement can easily be proved 
via the twistor argument. Namely, let us consider two rank~2 holomorphic vector
bundles over the minitwistor space~${\cal T}=T\C P^1$ with holomorphic
transition functions $f_{+-}$ and $\tilde f_{+-}$
(with operator entries in the noncommutative case) defined on an open subset
of $V_+\cap V_-\subset{\cal T}$. These bundles are called equivalent if there
exist matrix-valued holomorphic functions $h_+$ and $h_-$ on $V_+$ and $V_-$, 
respectively, such that
\begin{equation}
h_+\,f_{+-}\,h_- \= \tilde f_{+-}\ .
\end{equation}
These transition functions define the same configuration since
the splitting of~$\tilde f_{+-}$ yields $\tilde\psi_+=\psi_+ h_+^{-1}$ and
$\tilde\psi_-=\psi_- h_-$, and the factors~$h_{\pm}$ disappear from 
the equations due to holomorphicity.

After these remarks, we explicitly consider the transition function 
(\ref{fSU2n}) and take
\begin{align}
h_+ &\=
\begin{pmatrix}
\prod\limits_{k=1}^n(\eta{-}h_k) &1\\ 1 & 0
\end{pmatrix} \=
\begin{pmatrix}
\la^n\r_n &1\\[6pt] 1 & 0
\end{pmatrix}\ , \nonumber\\[8pt]
h_- &\=
\begin{pmatrix}
\!{-}\!\prod\limits_{k=1}^n\bigl(\te{-}h_k (\lt )\bigr) &1\\ 1 & 0
\end{pmatrix} \=
\begin{pmatrix}
\!{-}\la^{-2n}\prod\limits_{k=1}^n\bigl(\h{-}h_k (\la)\bigr) &1\\ 1 & 0
\end{pmatrix} \=
\begin{pmatrix}
\!{-}\la^{-n}\r^{-1}_n &1\\[6pt] 1 & 0
\end{pmatrix}\ ,
\end{align}
which are holomorphic on $V_+$ and $V_-$, respectively. Then, we have
\begin{align}\label{fWYn}
h_+\,f_{+-}^{(n)WY} h_- &\=
\begin{pmatrix}
\la^{n}\r^{-1}_n &1\\[6pt] 1 & 0
\end{pmatrix}
\begin{pmatrix}
\r_n& \la^{-n}\\[8pt] -\la^{n} & 0
\end{pmatrix}
\begin{pmatrix}
-\la^{-n}\r^{-1}_n\ &1\\[6pt] 1 & 0
\end{pmatrix} \nonumber\\[8pt]
&\= \begin{pmatrix}
\r^{-1}_n & 0 \\[6pt] 0 & \r_n
\end{pmatrix} \=
\begin{pmatrix}
(f_{+-}^{(n)D})^{-1} & 0 \\ 0 & f_{+-}^{(n)D}
\end{pmatrix}\ ,
\end{align}
where $f_{+-}^{(n)D}$ defines the Dirac multi-monopole,
and the diagonal matrix~(\ref{fWYn}) describes the line bundle~$L$ 
(the U(1) gauge group) embedded into the rank~2 vector bundle 
(the SU(2) gauge group) as $L^{-1}\oplus L$.
This proves the equivalence. The same formulae can be written in the
noncommutative case since all calculations above are purely algebraic.

\end{appendix}

\end{document}